\documentclass[aps,prd,reprint,amsmath,amssymb]{revtex4-1}
\usepackage{graphicx}
\usepackage{dcolumn}
\usepackage{bm}
\usepackage{color}

\newcommand{\RN}[1]{%
  \textup{\uppercase\expandafter{\romannumeral#1}}%
}

\begin{document}

\preprint{}

\title{High-energy cosmic ray nuclei from tidal disruption events: \\
Origin, survival, and implications}



\author{B. Theodore Zhang$^{1,2,3}$}
\author{Kohta Murase$^{3,4,5,6}$}
\author{Foteini Oikonomou$^{3, 5}$}
\author{Zhuo Li$^{1,2}$}

\affiliation{$^1$Department of Astronomy, School of Physics, Peking University, Beijing 100871, China }
\affiliation{$^2$Kavli Institute for Astronomy and Astrophysics, Peking University, Beijing 100871, China}
\affiliation{$^3$Department of Physics, The Pennsylvania State University, University Park, Pennsylvania 16802, USA}
\affiliation{$^4$Department of Astronomy \& Astrophysics, The Pennsylvania State University, University Park, Pennsylvania 16802, USA}
\affiliation{$^5$Center for Particle and Gravitational Astrophysics, The Pennsylvania State University, University Park, Pennsylvania 16802, USA}
\affiliation{$^6$Yukawa Institute for Theoretical Physics, Kyoto, Kyoto 606-8502 Japan}

\date{\today}

\begin{abstract}
Tidal disruption events (TDEs) by supermassive or intermediate mass black holes have been suggested as candidate sources of ultrahigh-energy cosmic rays (UHECRs) and high-energy neutrinos. Motivated by the recent measurements from the Pierre Auger Observatory, which indicates a metal-rich cosmic-ray composition at ultrahigh energies, we investigate the fate of UHECR nuclei loaded in TDE jets. First, we consider the production and survival of UHECR nuclei at internal shocks, external forward and reverse shocks, and nonrelativistic winds. Based on the observations of Swift J1644+57, we show that the UHECRs can survive for external reverse and forward shocks, and disk winds. On the other hand, UHECR nuclei are significantly disintegrated in internal shocks, although they could survive for low-luminosity TDE jets.  Assuming that UHECR nuclei can survive, we consider implications of different composition models of TDEs. We find that the tidal disruption of main sequence stars or carbon-oxygen white dwarfs does not successfully reproduce UHECR observations, namely the observed composition or spectrum. The observed mean depth of the shower maximum and its deviation could be explained by oxygen-neon-magnesium white dwarfs, although they may be too rare to be the sources of UHECRs. 

\end{abstract}

\maketitle


\section{\label{sec:level1}Introduction}

Cosmic rays with energy larger than $10^{18}~\rm eV$ are referred to as ultrahigh-energy cosmic rays (UHECRs), and their origin is still largely unknown \cite{Hillas:1985is, Nagano:2000ve, Kotera:2011cp}. The observed spectrum of UHECRs has a cutoff at energy around $\sim 4 \times 10^{19} \rm~eV$ \cite{Abraham:2008ru, Abbasi:2007sv, Abraham:2010mj}. The flux suppression is consistent with the prediction of the Greisen-Zatsepin-Kuzmin (GZK) effect due to the interaction between UHECRs and cosmic microwave background (CMB) photons \cite{Greisen:1966jv, Zatsepin:1966jv}. A key clue to the origin of UHECRs is their composition. The primary mass of UHECRs can be inferred from the distributions of the depth of the shower maximum, $X_{\rm max}$. The data from the Telescope Array (TA) and Auger are consistent within systematic and statistical uncertainties \cite{Abbasi:2015xga}. The analysis from Auger suggests that the composition of UHECRs is dominated by light nuclei at energy around $10^{18.3}\rm~eV$ and becomes gradually heavier with increasing energy up to $10^{19.6}\rm~eV$ \cite{Aab:2014kda, Aab:2014aea}. The distributions of $X_{\rm max}$ are difficult to explain with a mixture of protons and iron nuclei over the whole energy range if up-to-date hadronic interaction models are correct. Rather, the best fit is reached by including a fraction of intermediate mass nuclei \cite{Aab:2014aea}. 

The interpretation of the UHECR composition is still under intense debate, and we avoid discussion that depends on their details. However, considering the larger detection area and lower sampling bias of Auger, it is reasonable to assume that UHECR composition gets gradually heavier at the highest energies. There are not so many candidate sources which satisfy the Hillas criterion to accelerate cosmic rays (CRs) to ultrahigh energies \cite{Hillas:1985is}, and the origin of heavy nuclei has now become an interesting question. 
In relativistic jets of active galactic nuclei (AGN) (e.g.,~\cite{Norman1995, Peer:2009vnw,dermer:2010iz,Murase:2011cy,Fang:2017zjf} and references therein), heavy nuclei would be supplied by an accretion disk.  For structure formation shocks in galaxy clusters (GCs) \cite{Norman1995, Kang:1995xw, Inoue:2007kn}, heavy nuclei can be provided from the intracluster medium. However, for these objects, we typically expect a composition similar to the solar composition, unless reacceleration is invoked \cite{Kimura:2017ubz, Caprioli:2015zka}. Another possibility is that heavy nuclei can be synthesized in the stellar interiors or outflows from the deaths of massive stars. This scenario is appropriate for gamma-ray bursts (GRBs) \cite{Murase:2008mr, Wang:2007xj, Horiuchi:2012by, Globus:2014fka, Biehl:2017zlw} (see \cite{Waxman:1995vg, Vietri:1995hs} for protons), low-luminosity GRBs and transrelativistic supernovae \cite{Murase:2006mm,Murase:2008mr,Wang:2007ya,Chakraborti:2010ha,Liu:2011tv}, and newborn pulsars and magnetars \cite{Fang:2012rx, Fang:2013vla, Kotera:2015pya}. However, UHECR nuclei can be depleted before they escape the source environment due to the interaction with background particles or photons, so the ``UHECR survival problem'' is important to discuss the origin of UHECR nuclei~\cite{Murase:2008mr, Wang:2007xj, Peer:2009vnw, Horiuchi:2012by,Kotera:2015pya}.

In this work, we revisit tidal disruption events (TDEs) as sources of UHECR nuclei. 
A TDE is a luminous flare lasting for months to years, which occurs in a galaxy nuclear region \cite{Rees:1988bf}. When a star gets very close to the central black hole, it is disrupted if the tidal force is greater than the star's self-gravity. During the disruption process, nearly half of the stellar debris falls back into the vicinity of the black hole, and the rest becomes unbound from the system. The accretion flow undertakes a fast energy dissipation and circularization process, and then it forms an accretion disk around the central black hole \cite{Shen:2013oma, Piran:2015gha}. The TDE may be accompanied by the emergence of a relativistic jet during the super-Eddington accretion phase \cite{Giannios:2011it, Lei:2011qg, Krolik:2011fb, DeColle:2012np, Shcherbakov:2012zt, McKinney:2013txa, MacLeod:2014mha}, and TDEs have been proposed to accelerate particles to ultrahigh energies \cite{Farrar:2008ex, Farrar:2014yla,Pfeffer:2015idq}.
Two scenarios are the most popular: disruption of a main-sequence (MS) star by a supermassive black hole (SMBH) and the disruption of a white dwarf (WD) by an intermediate mass black hole (IMBH). The latter could be especially interesting as sources that can inject heavy nuclei.  

This paper is organized as follows. In Sec.~\ref{sec:two}, we show that both protons and nuclei can be accelerated to ultrahigh energies, and we study conditions for the survival of UHECR nuclei in jetted TDEs. For the acceleration sites, we consider internal shocks, external reverse and forward shocks, and nonrelativistic winds. In Sec.~\ref{sec:three}, for different composition models, we study the propagation of UHECR nuclei using the public code CRPropa 3 \cite{Batista:2016yrx}, and compare the results to experimental results by Auger. 

Throughout the paper, we use cgs units, and adopt notations such as $Q_x \equiv Q /10^x$. The cosmological parameters we assume are $H_0 = 67.3 \ \rm km \ s^{-1} \ Mpc^{-1}$, $\Omega_m = 0.315$, and $\Omega_\Lambda = 0.685$ \cite{Agashe:2014kda}.

\section{\label{sec:two}Survival of cosmic-ray nuclei in TDE Outflows}
We discuss possible composition models in the next section. In this section, we consider the fate of UHECR nuclei, following Ref.~\cite{Murase:2008mr}. 

Jetted TDEs (e.g., Swift J1644+57) show clear signatures of nonthermal emission in a wide range of wavelengths from radio to x rays \cite{Bloom:2011xk, Burrows:2011dn, Zauderer:2011mf}. 
Diffusive shock acceleration in collisionless shocks is the most popular mechanism of production of nonthermal particles. 
The first-order Fermi process is predicted to have a power-law distribution of accelerated particles within the ``test particle'' approximation. The acceleration time scale can be expressed as $t_{\rm acc} = \eta r_L/c$, where $r_L = E_A / Z e B$ is the Larmor radius of a particle with energy $E_A$, charge number $Z$, and $B$ the comoving-frame magnetic field strength. The factor $\eta$ depends on details of the turbulence. The minimum value of $\eta \sim 1-10$ can be achieved in the Bohm limit \cite{Lemoine2009}, and $\eta=1$ is used to demonstrate our results in this section.

The maximum acceleration energy is determined by $t_{\rm acc} \leq {\rm min}(t_{\rm dyn}, t_{\rm syn}, t_{A\gamma})$, where $t_{\rm dyn} \equiv R/ \Gamma \beta c$ is the dynamical time scale; $t_{\rm syn} = 3m_A^4 c^3 \Gamma /(4 \sigma_T Z^4 m_e^2 U_B E_A)$ is the synchrotron cooling time scale ($\sigma_T$ is the Thompson cross-section, $U_B = B^2/8\pi$ is the magnetic energy density, $E_A$ is the particle energy); and $t_{A\gamma}$ is the energy loss time scale for protons (photomeson interaction) and nuclei (photodisintegration interaction). 
We can estimate the energy loss time scale using the following formula,
\begin{equation}
t_{A\gamma}^{-1}(E_A) = \frac{c}{2 \gamma_A^2} \int_{\bar{\varepsilon}_{\rm th}}^{\infty} 
d\bar{\varepsilon} \sigma_{A\gamma}(\bar{\varepsilon})  \kappa_{A\gamma}(\bar{\varepsilon})\bar{\varepsilon}\int_{\bar{\varepsilon}/2\gamma_A}^{\infty} d\varepsilon \frac{1}{\varepsilon^2} \frac{dn}{d\varepsilon},
\end{equation}
where $\gamma_A$ is the Lorentz factor of UHECRs with mass number $A$, $\bar{\varepsilon}_{\rm th}$ is the threshold energy measured in the rest frame of initial nucleus (NRF), and ${\rm d} n/{\rm d}\varepsilon$ is the differential number density of target photons. Here, $\sigma_{A\gamma}$ is the photohadronic cross section related to the photomeson or photodisintegration process, and $\kappa_{A\gamma}$ is the inelasticity of each process.  We show photonuclear and photohadronic cross sections for five typical chemical species as a function of NRF target photon energy ranging from $\sim 10 \ \rm MeV$ to $\sim 10^7 \ \rm GeV$  in Fig.~\ref{fig:crosssection}; see Appendix~\ref{Appendix_A}. 

\begin{figure}
\includegraphics[width=\linewidth]{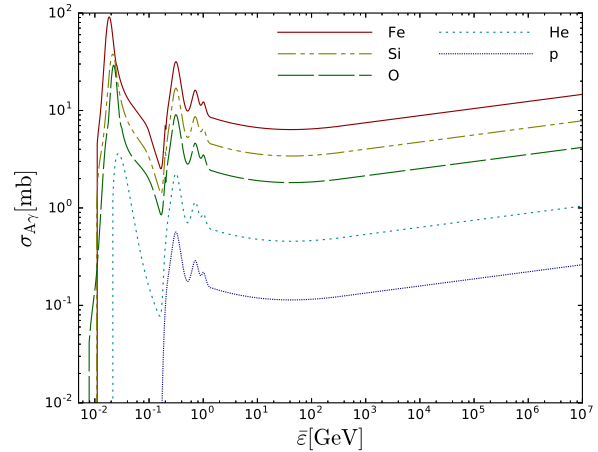}
\caption{Photonuclear and photomeson production cross sections for five chemical species, which are used in this work: Fe, Si, O, He, and proton, as a function of NRF target photon energy \cite{Rachen1996, Agostinelli:2002hh, Koning2007, Batista:2016yrx}. For simplicity, the superposition model is assumed for the photomeson production. \label{fig:crosssection}}
\end{figure}

Defining the optical depth as $f_{A\gamma} \equiv t_{A\gamma}^{-1} / t_{\rm dyn}^{-1}$, we can expect the survival of UHECR nuclei when $f_{A\gamma} < 1$. The value of the interaction time scale $t_{A\gamma-\rm int}$ is equal to the energy loss time scale when $\kappa_{A\gamma} = 1$, and the related optical depth is $\tau_{A\gamma} \equiv t_{A\gamma-\rm int}^{-1} / t_{\rm dyn}^{-1}$.

\subsection{Internal shock model}
Nonthermal hard x-ray emission comes from internal energy dissipation in the inner relativistic jet \cite{Burrows:2011dn}. 
In the internal shock region, fast moving ejecta may catch up with slower ejecta, and a substantial amount of kinetic energy of the relativistic jet may be converted into internal energy. We assume the radius where internal collisions take place is $R_{\rm IS} =  \Gamma^2 c \delta t = 3 \times 10^{14} \ \Gamma_1^2 \delta t_2  \ \rm cm$, with $\Gamma = 10 \Gamma_1$ the Lorentz factor of the relativistic jet and $\delta t \sim 100 \delta t_2 \rm \ s$ is the x-ray variability time scale \cite{Burrows:2011dn}. The median x-ray luminosity of Swift J1644+57 is $L_{\rm X, iso} = 8.5 \times 10^{46} \ \rm{erg \ s^{-1}}$, which is well above the Eddington luminosity $L_{\rm Edd} = 1.3 \times 10^{44} M_{\rm BH, 6} \ \rm erg \ s^{-1}$ of a $10^6 M_\odot$ black hole \cite{Burrows:2011dn}. In our analysis, the total isotropic luminosity is set to $L_{\rm iso} = 10^{48} \ \rm erg \ s^{-1}$. A fraction $\epsilon_B$ of the total energy of the outflow is converted into magnetic energy $B^2/8\pi = \epsilon_B L_{\rm iso}/4\pi R_{\rm IS}^2 \Gamma^2 c$. The magnetic field strength in the jet comoving frame is $B= (2 \epsilon_B L_{\rm iso} / R_{\rm IS}^2 \Gamma^2 c)^{1/2}\simeq 860 \ L_{\rm iso, 48}^{1/2} \epsilon_{B,-1}^{1/2} R_{\rm IS, 14.5}^{-1} \Gamma_{1}^{-1} \rm~G $. The maximum acceleration energy can be achieved under the condition $t_{\rm acc} \leq t_{\rm dyn}$; we have 
$E_{A, \rm dyn} \simeq \Gamma \eta^{-1} Z e B (R_{\rm IS} / \Gamma) \sim 1.5 \times 10^{20}  \ Z \eta^{-1} L_{\rm iso, 48}^{1/2} \epsilon_{B,-1}^{1/2} \Gamma_{1}^{-1} \ \rm~eV$ measured in the observer frame, where $R_{\rm IS} / \Gamma$ is the comoving frame shell width. The maximum acceleration energy is also limited by the synchrotron energy loss $t_{\rm acc} < t_{\rm syn}$; we have $E_{A, \rm syn} \sim 6.3 \times 10^{19} \ A^2 Z^{-3/2} \eta^{-1/2} L_{\rm iso,48}^{-1/4} \epsilon_{B, -1}^{-1/4} \Gamma_{1}^{3/2} R_{\rm IS, 14.5}^{1/2} \rm~eV$ measured in the observer frame. To estimate the effect of photonuclear and photohadronic interactions, we use a log-parabola model to fit the high-luminosity state SED of Swift J1644+57 \cite{Burrows:2011dn, Senno:2016bso}.

\begin{figure}
\includegraphics[width=\linewidth]{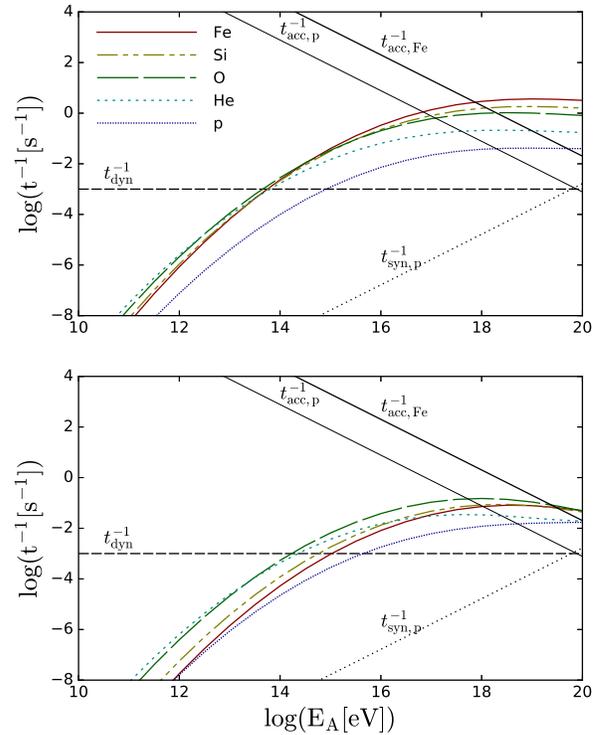}
\caption{Various time scales for five typical chemical species (Fe, Si, O, He, and proton) in the internal shock region as a function of particle energy (measured in the observer frame). We show interaction time scales in the upper panel and energy-loss time scales in the lower panel. The thin (thick) black line represents the acceleration time scale for the proton (Fe). We show the synchrotron cooling time scale for protons as the dotted black line. The parameters are $L_{\rm iso} = 10^{48} \rm \ erg \ s^{-1}$, $\Gamma = 10$, $\epsilon_B = 0.1$, and $r = 3 \times 10^{14} \rm \ cm$. \label{fig:IS_timescale}}
\end{figure}

Our results for the internal shock scenario are shown in Fig.~\ref{fig:IS_timescale}. We consider two time scales, one is the interaction time scale, $t_{A\gamma-\rm int}$, and the other is the energy-loss time scale, $t_{A\gamma}$. We see that $t_{A\gamma-\rm int}$ and $t_{A\gamma}$ are shorter than $t_{\rm dyn}$. 
Our calculation suggests that most CR nuclei in the internal shock region will be disintegrated before they escape, so it is difficult for CR nuclei to survive for luminous TDEs like Swift J1644+57. Note that even partial survival is difficult for this parameter set (i.e., $f_{A\gamma}\gtrsim1$), implying that the composition becomes light due to efficient photodisintegration. 

\begin{figure}
\includegraphics[width=\linewidth]{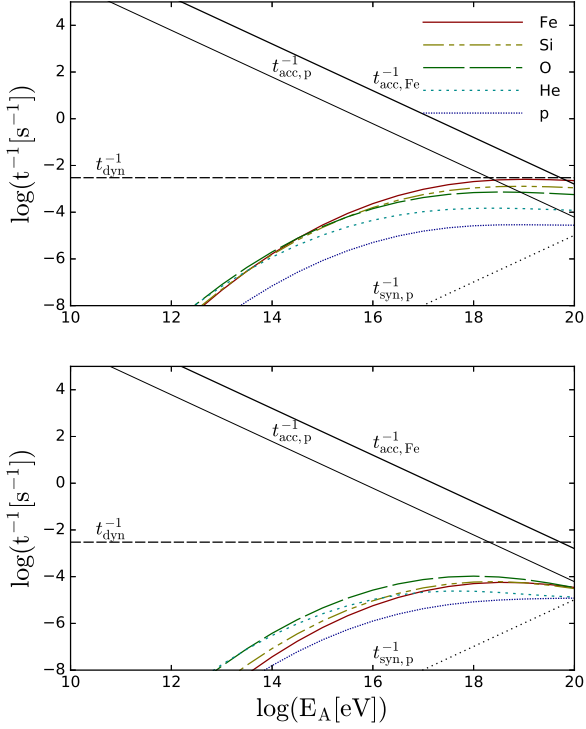}
\caption{Same as Fig.~\ref{fig:IS_timescale}, but the parameters are $L_{\rm iso} = 10^{44.8} \rm \ erg \ s^{-1}$, $\Gamma = 10$, $\epsilon_B = 0.1$, and $r = 10^{14} \rm \ cm$. \label{fig:IS_L_timescale}}
\end{figure}

We also consider the survival of UHECR nuclei in TDEs with a lower luminosity, and we assume an isotropic luminosity of $L_{\rm iso} = 10^{44.8}\ \rm erg \ s^{-1}$. The comoving frame magnetic field strength is estimated to be $B \simeq 96 \ L_{\rm iso, 44.8}^{1/2} \epsilon_{B, -1}^{1/2} R_{\rm IS, 14}^{-1} \Gamma_1^{-1} \rm~G$. The results are shown in Fig.~\ref{fig:IS_L_timescale}, and we find that UHECR nuclei with energy up to $\sim 10^{20} \ \rm eV$ can survive before they escape from the source. Note that this case almost allows the complete survival of nuclei (i.e., $\tau_{A\gamma}\lesssim1$). For intermediate luminosities corresponding to $\tau_{A\gamma}\gtrsim1$ and $f_{A\gamma}\lesssim1$, UHECR nuclei partially survive~\cite{Murase:2010va} and secondary nuclei affect the initial composition of UHECRs. 

\subsection{Reverse shock model}
The radio afterglow of Swift J1644+57 was observed a few days after the trigger of the Swift BAT observation \cite{Zauderer:2011mf}, and the continued observation extends to $\sim 500$ days \cite{Berger:2011aa, Zauderer:2012mr}. The nonthermal radio emission is consistent with synchrotron radiation from the standard external shock model \cite{Metzger:2011qq, Zauderer:2012mr}. Here, we assume the jet duration time is $t_j = 10^6\ \rm s$ based on the observation of Swift J1644+57 \cite{Burrows:2011dn}. The isotropic equivalent kinetic energy of the jet is $L_{\rm iso} \sim 10^{47} \ \rm erg \ s^{-1}$ on average, and it will decrease following $L_{\rm iso} \propto t^{-5/3}$ after the time $t_j$. We estimate the total injection energy as ${\mathcal E}_{\rm iso} = 2 L_{\rm iso} t_j = 2 \times 10^{53} \ \rm erg$. The relativistic jet is decelerated when it has swept up a significant amount of circumnuclear medium (CNM). There are two shocks formed in the deceleration radius, one is the reverse shock propagating back into the relativistic jet, and the other is the forward shock propagating into the CNM. For simplicity, we adopt a constant CNM density, $\varrho = 1 \ \rm cm^{-3}$. 

We calculate the evolution of reverse shock following the same method as the one used for GRBs \cite{Panaitescu:2004sn, Murase:2007yt, Murase:2008mr}. The Lorentz factor of the shocked ejecta (relative to unshocked CNM) can be estimated as $\Gamma \approx \Gamma_0 / (1+2\Gamma_0(\varrho / n_j)^{1/2})^{1/2}$ assuming the pressure equilibrium at the contact discontinuity. We assume $\Gamma_0 = 10$, and the density of the jet $n_j = {\mathcal E}_{\rm iso} / (4\pi m_p c^2 \Gamma_0 (\Gamma_0 \Delta) r^2)$, where ${\mathcal E}_{\rm iso}$ is the isotropic equivalent energy of the relativistic outflow, and $\Delta$ is the thickness of the ejecta (relative to the central black hole). The thickness of the ejecta is estimated to be $\Delta_\times \approx \Delta_0 \equiv c t_j$. The shock completely crosses the ejecta at $t = t_\times$. We write the radius at that time as $r_\times \simeq 8.9 \times 10^{17} {\mathcal E}_{\rm iso, 53.3}^{1/4} t_{j, 6}^{1/4} \varrho^{-1/4} \ {\rm cm}$ and the Lorentz factor at the crossing time is $\Gamma_\times \simeq 3.2 {\mathcal E}_{\rm iso, 53.3}^{1/8} t_{j, 6}^{-3/8} \varrho^{-1/8}$. The magnetic field strength is estimated to be $B_\times =(32 \pi \epsilon_B^r n_j m_p c^2(\Gamma_\times -1)(\Gamma_\times + 3/4))^{1/2}$, where a fraction $\epsilon_B^r$ of the post-shock internal energy is converted into magnetic energy. The comoving frame Lorentz factor of the electrons is $\gamma_e^r = \frac{\epsilon_e^r}{f_e^r}\frac{p-2}{p-1}\frac{m_p}{m_e}(\Gamma_\times - 1)$, with $p = 2.2$, $\epsilon_e^r = 0.02$, and $f_e^r = 0.1$. Here $f_e^r$ represents the number fraction of accelerated electrons. 

Then, the peak synchrotron frequency is written as
\begin{eqnarray}
\nu_{m, \rm{ob}}^r \simeq 8 &\times& 10^{10} \frac{g_1(\Gamma_\times)}{g_1(3.2)} {\mathcal E}_{\rm iso, 53.3}^{1/4} \nonumber \\ &\times& {\epsilon_{B, -2}^r}^{1/2} {\epsilon_{e, -1.7}^r}^2 {f_{e, -1}^r}^{-2} \varrho^{1/4} t_{j, 6}^{-3/4}  \ \rm~Hz,
\end{eqnarray}
where $g_1(\Gamma_\times) = \Gamma_\times (\Gamma_\times - 1)^{5/2}(\Gamma_\times + 3/4)^{1/2}$. The synchrotron self-absorption frequency can be estimated when $\tau(\nu_a) = 1$; we have ($\nu_a < \nu_m$)
\begin{eqnarray}
\nu_{a, \rm{ob}}^r \simeq  5.6 &\times& 10^{9} \frac{g_2(\Gamma_\times)}{g_2(3.2)} {\mathcal E}_{\rm iso, 53.3}^{8/20} {\epsilon_{B, -2}^r}^{1/5} \nonumber \\ &\times&  {\epsilon_{e, -1.7}^r}^{-1} {f_{e, -1}^r}^{8/5} \varrho^{8/20} t_{j, 6}^{-3/5}\rm~Hz,
\end{eqnarray}
 where $g_2(\Gamma_\times) = g_1(\Gamma_\times) (\Gamma_\times - 1)^{-33/10} (\Gamma_\times + 3/4)^{-3/10}$. 
Further, we can express the comoving frame peak luminosity per unit energy from the reverse shock as
\begin{eqnarray}
L_{\epsilon, \rm max}^r &=& \frac{1}{2\pi\hbar} N_e^r f_e^r \frac{\sqrt{3} e^3 B_\times}{m_e c^2} \nonumber \\ 
&\simeq& 1.14 \times 10^{58} \ \frac{g_3(\Gamma_\times)}{g_3(3.2)} {\mathcal E}_{\rm iso, 53.3}^{5/4} \nonumber \\ &\times& {\epsilon_{B, -2}^r}^{1/2} {f_{e, -1}^r} \varrho^{1/4} t_{j, 6}^{-3/4} \rm~s^{-1},
\end{eqnarray}
where $g_3(\Gamma_\times) = (\Gamma_\times - 1)^{1/2}(\Gamma_\times + 3/4)^{1/2}$ and $N_e^r = {\mathcal E}_{\rm iso}/ \Gamma_0 m_p c^2$ is the number of electrons in the reverse shock. The photon spectrum in the reverse shock is ($\nu_a < \nu_m < \nu_c$) 
\begin{equation}
\frac{dn^r}{d\varepsilon} = \frac{L_{\varepsilon, \rm max}^r}{4 \pi r_\times^2 c \varepsilon_m} \left\{ \begin{array}{ll} (\frac{\varepsilon_a}{\varepsilon_m})^{-2/3}(\frac{\varepsilon}{\varepsilon_a})^{1} & (\varepsilon < \varepsilon_a) \\ (\frac{\varepsilon}{\varepsilon_m})^{-2/3} & (\varepsilon_a \leq \varepsilon < \varepsilon_m) \\ (\frac{\varepsilon}{\varepsilon_m})^{-(p-1)/2 - 1} & (\varepsilon_m \leq \varepsilon < \varepsilon_c) \\ (\frac{\varepsilon_c}{\varepsilon_m})^{-(p-1)/2 - 1} (\frac{\varepsilon}{\varepsilon_c})^{-p/2 - 1} & (\varepsilon \geq \varepsilon_c) \end{array} \right.,
\end{equation}
where $\varepsilon_m = \varepsilon_{m, {\rm ob}} / \Gamma_\times$ is the peak photon energy in the reverse shock comoving frame and $\varepsilon_c$ is the electron synchrotron cooling energy.
The maximum acceleration energy of nuclei can be derived when $t_{\rm acc} = t_{\rm dyn}$, 
\begin{eqnarray}
E_{A,\rm max} &=& \Gamma_\times \eta^{-1} Z e B_\times (r_\times / \Gamma_\times) \nonumber\\
&\simeq& 6.5 \times 10^{19} Z \eta^{-1} \frac{g_3(\Gamma_\times)}{g_3(3.2)}  \nonumber \\ 
&\times& {\mathcal E}_{\rm iso, 53.3}^{1/2} {\epsilon_{B,-2}^r}^{1/2} t_{j, 6}^{-1/2} \rm~eV.
\end{eqnarray}

\begin{figure}
\includegraphics[width=\linewidth]{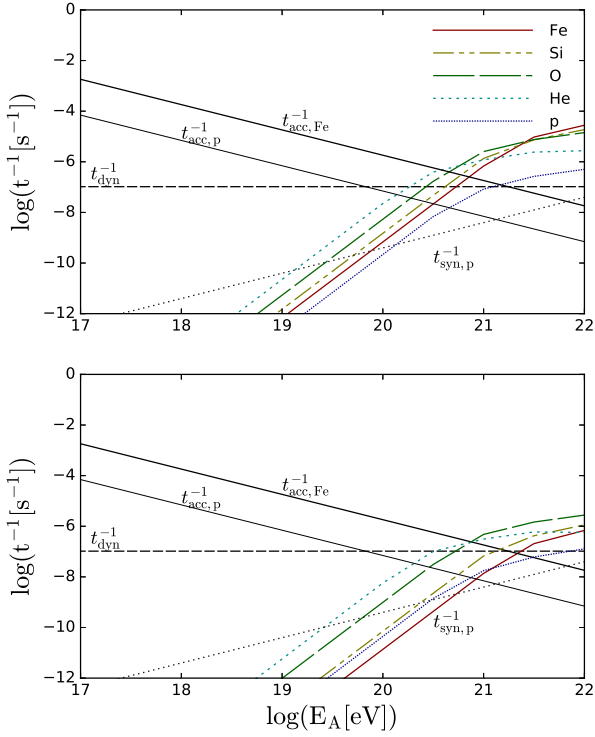}
\caption{Various time scales for five typical chemical species (Fe, Si, O, He, and proton) in the reverse shock region as a function of particle energy (measured in the observer frame). We show the interaction time scales in the upper panel and energy-loss time scales in the lower panel. The thin (thick) black line represents the acceleration time scale for the proton (Fe). We show the synchrotron cooling time scale for the proton as the dotted black line. The parameters we use are $L_{\rm iso} = 10^{47} \rm \ erg \ s^{-1}$, $t_j = 10^6 \ \rm s$, $\Gamma_0 = 10$, $\epsilon_B^r = 0.01$, $\epsilon_e^r = 0.02$, and $f_e^r = 0.1$.\label{fig:RS_timescale}}
\end{figure}

We estimate various time scales in the reverse shock model, and our results are shown in Fig.~\ref{fig:RS_timescale}. We find that the interaction time scale $t_{A\gamma-\rm int}$ is longer than the dynamical time scale $t_{\rm dyn}$, which means that the optical depth $\tau_{A\gamma} < 1$. Our results suggest that UHECR nuclei can survive in the reverse shock model. 

\subsection{Forward shock model}
Once the relativistic jet enters the CNM, the deceleration and transition to the Blandford-Mckee (BM) self-similar regime occurs. The evolution of the Lorentz factor and shock radius are described as $\Gamma(t) \propto t^{-3/8}$ and $r(t) \propto t^{1/4}$, where we assume a constant CNM density. The evolution of the downstream magnetic field follows $B \propto t^{-3/8}$. A fraction of electrons are accelerated in the shock, with a minimum Lorentz factor, $\gamma_m$, and the distribution of accelerated electrons is denoted as $dN_e/d\gamma_e \propto \gamma^{-p}$ with $p = 2.2$, and minimum Lorentz factor $\gamma_m \propto t^{-3/8}$. In the following, we use the approximation $\Gamma \gg 1$ to derive the simplified expression of typical frequency and luminosity, and we adopt the accurate formula in our calculation due to the mildly relativistic nature of forward shock. The peak frequency from the forward shock can be estimated as 
\begin{equation}
\nu_{m, {\rm ob}} \simeq  1.3 \times 10^{11} \ {\mathcal E}_{\rm iso, 53.3}^{1/2} \epsilon_{B, -1.8}^{1/2} \epsilon_{e, -1.5}^2 f_{e, -1}^{-2}  t_{6}^{-3/2} \ \rm Hz ,
\end{equation}
The self-absorption frequency (assuming $\nu_a < \nu_m < \nu_c$) is
\begin{equation}
\nu_{a, {\rm ob}} \simeq 3.1 \times 10^{9} {\mathcal E}_{\rm iso, 53.3}^{1/5} \epsilon_{B, -1.8}^{1/5} \epsilon_{e, -1.5}^{-1} f_{e,-1}^{3/5} \varrho^{3/5} \ \rm Hz.
\end{equation}

The comoving frame peak luminosity per unit energy is 
\begin{eqnarray}
L_{\epsilon, \rm{max}} &=& \frac{1}{2\pi \hbar}N_e f_e \frac{\sqrt{3} e^3 B}{m_e c^2}  \nonumber \\ 
&\simeq& 9.8 \times 10^{57} {\mathcal E}_{\rm iso, 53.3}^{7/8} \epsilon_{B, -1.8}^{1/2} f_{e, -1} \varrho^{5/8} t_{6}^{3/8} \ \rm s^{-1},
\end{eqnarray}
where $N_e = (4\pi/3) r^3 \varrho$ is the total number of swept-up electrons. The comoving frame photon spectrum can be derived in the slow cooling case ($\nu_a < \nu_m < \nu_c$),
\begin{equation}
\frac{dn^f}{d\varepsilon} = \frac{L_{\varepsilon, \rm max}}{4 \pi r^2 c \varepsilon_m} \left\{ \begin{array}{ll} (\frac{\varepsilon_a}{\varepsilon_m})^{-2/3}(\frac{\varepsilon}{\varepsilon_a})^{1} & (\varepsilon < \varepsilon_a) \\ (\frac{\varepsilon}{\varepsilon_m})^{-2/3} & (\varepsilon_a \leq \varepsilon < \varepsilon_m) \\ (\frac{\varepsilon}{\varepsilon_m})^{-(p-1)/2 - 1} & (\varepsilon_m \leq \varepsilon < \varepsilon_c) \\ (\frac{\varepsilon_c}{\varepsilon_m})^{-(p-1)/2 - 1} (\frac{\varepsilon}{\varepsilon_c})^{-p/2 - 1} & (\varepsilon \geq \varepsilon_c) \end{array} \right.,
\end{equation}
where $\varepsilon_m = \varepsilon_{m, {\rm ob}} / \Gamma$ is the peak photon energy in the forward shock comoving frame and $\varepsilon_c$ is the electron synchrotron cooling energy.

\begin{figure}
\includegraphics[width=\linewidth]{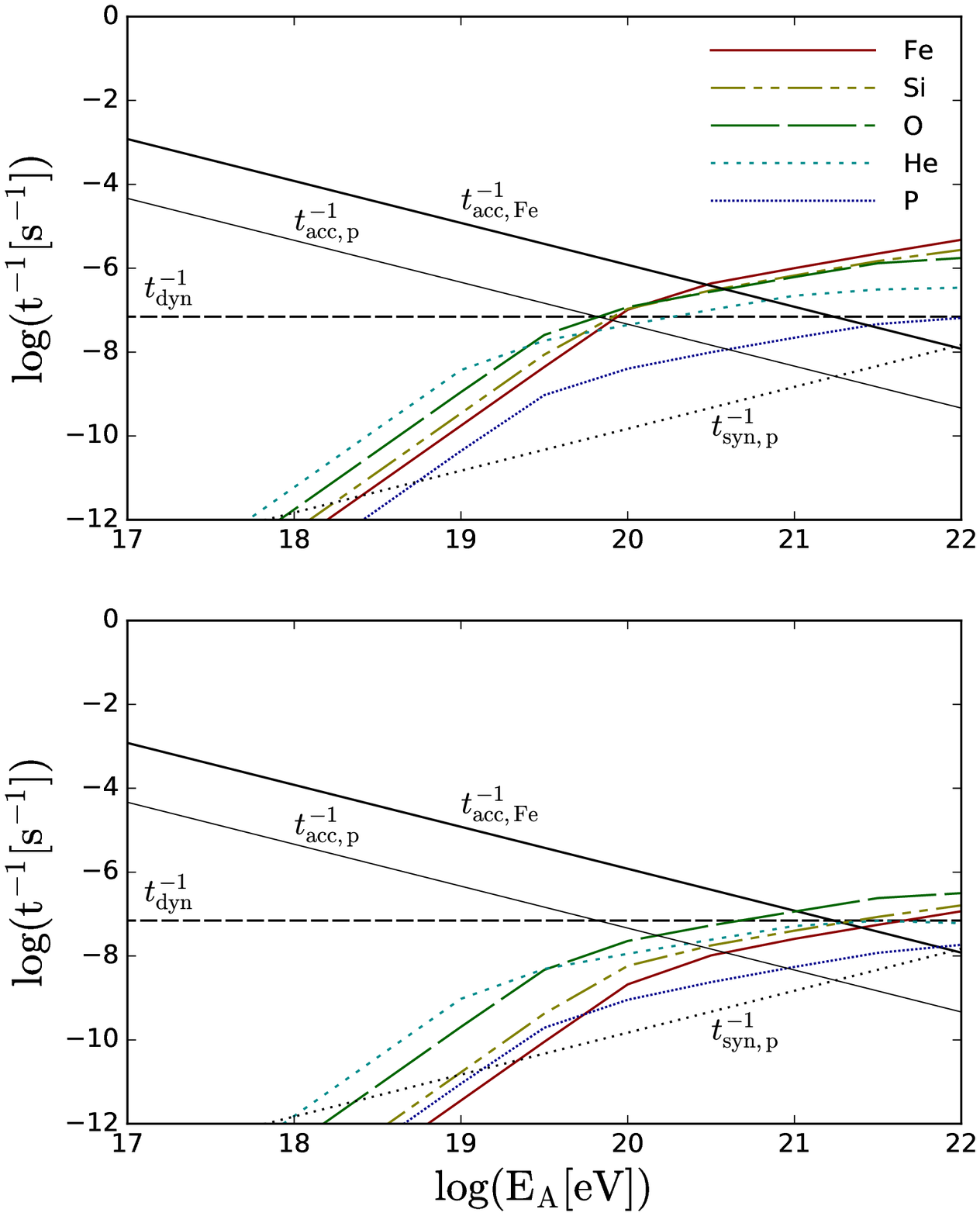}
\caption{Various time scales for five typical chemical species (Fe, Si, O, He, and proton) in the forward shock region as a function of particle energy (measured in the observer frame). We show the interaction time scales in the upper panel and energy-loss time scales in the lower panel. The thin (thick) black line represents the acceleration time scale for the proton (Fe). We show the synchrotron cooling time scale for the proton as the dotted black line. The parameters we use are $L_{\rm iso} = 10^{47} \rm \ erg \ s^{-1}$, $t = 10^6 \ \rm s$, $\Gamma_0 = 10$, $\epsilon_B = 0.015$, $\epsilon_e = 0.03$, and $f_e = 0.1$.\label{fig:FS_timescale}}
\end{figure}

\begin{figure}
\includegraphics[width=\linewidth]{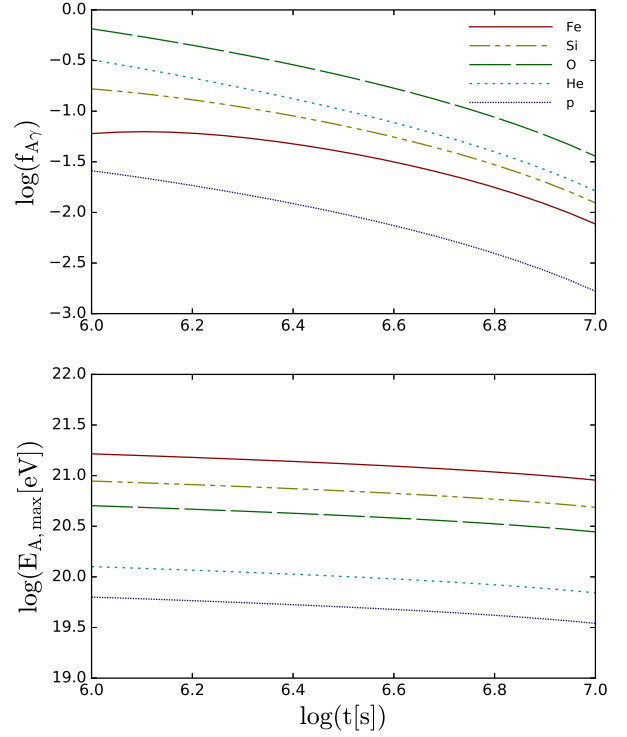}
\caption{The time evolution of maximum energy (lower panel) and optical depth $f_{A\gamma}$ (upper panel) for five typical chemical species: Fe, Si, O, He, and proton in the forward shock model. The optical depth is calculated when UHECR nuclei have energy $10^{20} \rm~eV$ (measured in the observer frame). \label{fig:FS_evolution}}
\end{figure}

In the forward shock model, the observed maximum particle energy is achieved when $t_{\rm acc} = t_{\rm dyn}$, with 
\begin{eqnarray}
E_{A,\rm max} &=& \Gamma \eta^{-1} Z e B (r / \Gamma) \\ \nonumber
&\simeq& 6.3 \times 10^{19} \ {\rm~eV} \ Z \eta^{-1} {\mathcal E}_{\rm iso, 53.3}^{3/8} \epsilon_{B, -1.8}^{1/2} \varrho^{1/8} t_{6}^{-1/8},
\end{eqnarray}
where it is assumed that the upstream magnetic field is amplified and comparable to the downstream value. The shock is mildly relativistic and this could be achieved by CR streaming instabilities \cite{Lemoine:2005zt, Lemoine:2006gg}. Or one could use the downstream field if particles are accelerated by the second-order Fermi acceleration mechanism \cite{Dermer:2000gu}.
We show various time scales in Fig.~\ref{fig:FS_timescale}. We expect the survival of UHECR nuclei at the forward shock, because the interaction time scale, $t_{A\gamma}$, is longer than the dynamical time scale, $t_{\rm dyn}$. We also show the evolution of particle maximum acceleration energy and optical depth $f_{A\gamma}$ in Fig.~\ref{fig:FS_evolution}. We found that it is easier for UHECR nuclei to survive at later times. 

However, it may be difficult to amplify the magnetic field in the upstream region, especially if the shock is ultrarelativistic \cite{ Murase:2007yt, Gallant:1998uq, Sironi:2015oza}. In this case, the magnetic field in the shock region should be similar to the CNM magnetic field $\hat{B}_{\rm cnm} = 10^{-5} \hat{B}_{\rm cnm, -5}$. The observed maximum acceleration energy is estimated to be
\begin{eqnarray}
E_{A, \rm max} &=& \Gamma \eta^{-1} Z e \Gamma \hat{B}_{\rm cnm} (r/\Gamma) \nonumber \\ &=&  4.2 \times 10^{15} \ {\rm~eV} \ Z {\mathcal E}_{\rm iso, 53.3}^{3/8} \varrho^{-3/8} \hat{B}_{\rm cnm, -5}t_{6}^{-1/8}.
\end{eqnarray}
It is difficult to accelerate CRs to ultrahigh energies if the magnetic field is not amplified efficiently.

\subsection{Nonrelativistic wind model}
Along with the formation of a relativistic jet, a nonrelativistic outflow can be driven by the accretion disk. The radio emissions from two TDEs, ASASSIN14li \cite{Alexander:2015jga} and XMMSL1J0740-85 \cite{Alexander:2016dxo}, are consistent with the emission from the interaction between a nonrelativistic outflow and the CNM.
We assume that the nonrelativistic outflow has an ejected mass of $M_{\rm ej} \sim 10^{-4} M_\odot$ and velocity $v_{\rm ej} \sim 0.1 c$. The deceleration radius is estimated to be $r_{\rm dec} \simeq 5.4 \times 10^{16} {\mathcal E}_{k, 48}^{1/3} \varrho^{-1/3} \ \rm cm$ and the deceleration time is $t_{\rm dec} \simeq 1.8 \times 10^{7}  {\mathcal E}_{k, 48}^{1/3} \varrho^{-1/3} \ \rm s$. The shocked fluid enters into the Sedov-Taylor evolution phase after it is decelerated by the external medium. In the adiabatic case, the shock velocity and radius follow $V_{\rm s} \propto t^{-3/5}$ and $R_{\rm s} \propto t^{2/5}$, respectively. We assume the accelerated electrons have power-law index $p = 3$.
The synchrotron peak frequency is 
\begin{equation}
\nu_m \simeq 1.5 \times 10^{6} {\mathcal E}_{k, 48} \epsilon_{B, -1}^{1/2} \epsilon_{e, -1}^2 f_{e, -1}^{-2} \varrho^{-1/2} (t/t_{\rm dec})^{-3} \rm \ Hz,
\end{equation}
and the self-absorption frequency ($\nu_a > \nu_m$) is 
\begin{equation}
\nu_a  \simeq 4.7 \times 10^7 \mathcal{E}_{k, 48}^{\frac{p}{p+4}} \epsilon_{B, -1}^{\frac{p/2+1}{p+4}} \epsilon_{e, -1}^{\frac{2p-2}{p+4}} f_{e, -1}^{\frac{2-2p}{p+4}} \varrho^{\frac{3-p/2}{p+4}} (t/t_{\rm dec})^{\frac{2-3p}{p+4}} \ \rm Hz,
\end{equation}
The photon spectrum in this case is($\nu_m < \nu_a < \nu_c$) 
\begin{equation}
\frac{dn}{d\varepsilon} = \frac{L_{\varepsilon, {\rm max}}}{4 \pi R_s^2 c \varepsilon_a} \left\{ \begin{array}{ll} (\frac{\varepsilon_m}{\varepsilon_a})^{5/2}(\frac{\varepsilon}{\varepsilon_a})^{-1}(\frac{\varepsilon}{\varepsilon_m})^{2} & (\varepsilon < \varepsilon_m) \\ (\frac{\varepsilon}{\varepsilon_a})^{3/2} & (\varepsilon_m \leq \varepsilon < \varepsilon_a) \\ (\frac{\varepsilon}{\varepsilon_a})^{-(p-1)/2- 1} & (\varepsilon_a \leq \varepsilon < \varepsilon_c) \\ (\frac{\varepsilon_c}{\varepsilon_a})^{-(p-1)/2- 1} (\frac{\varepsilon}{\varepsilon_c})^{-p/2-1} & (\varepsilon \geq \varepsilon_c) \end{array}\right..
\end{equation}

\begin{figure}
\includegraphics[width=\linewidth]{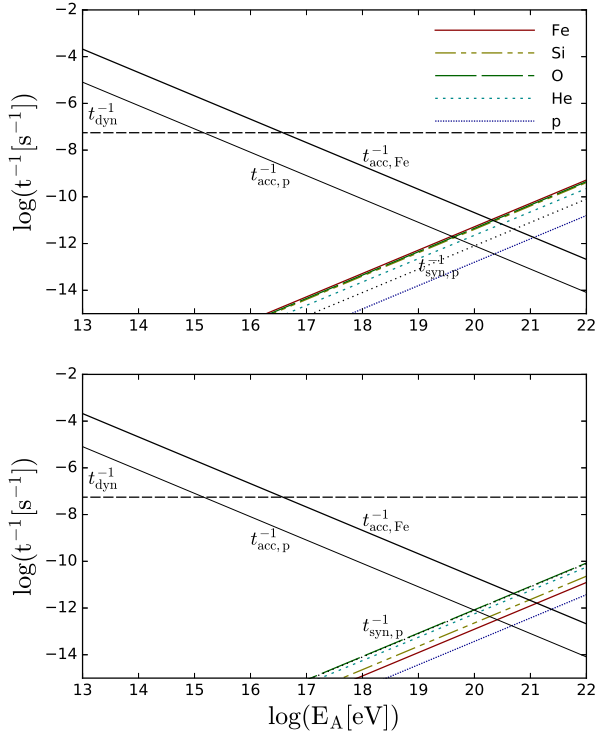}
\caption{Various time scales for five typical chemical species (Fe, Si, O, He, and proton) in the wind model as a function of particle energy. We show the interaction time scales in the upper panel and energy-loss time scales in the lower panel. The thin (thick) black line represents the acceleration time scale for the proton (Fe). We show the synchrotron cooling time scale for protons as the dotted black line.  The parameters we use are ${\mathcal E}_k = 10^{48} \rm \ erg $, $V_s = 0.1 c$, $\epsilon_B = 0.1$, $\epsilon_e = 0.1$, and $f_e = 0.1$.\label{fig:Wind_timescale}}
\end{figure}

In the nonrelativistic case, the particle acceleration time scale is  $t_{\rm acc} \approx (20/3) c^2 E_A / V_s^2 Z e B c$ in the Bohm limit. Our results for various time scales are shown in Fig ~\ref{fig:Wind_timescale}. The dynamical time in the nonrelativistic case is $t_{\rm dyn} = R_s / V_s$. The maximum acceleration energy is limited by $t_{\rm acc} \leq t_{\rm dyn}$, with $E_{A, \rm max} \approx (3/20) R_s Z e B V_s / c \sim 1.5 \times 10^{15} Z {\mathcal E}_{k, 48}^{3/5} \epsilon_{B, -1}^{1/2}  \varrho^{-1/10} (t/t_{\rm dec})^{-4/5} \ {\rm~eV}$. In the wind model, CRs cannot be accelerated to ultrahigh energies.

\section{\label{sec:three}Propagation of Escaping Nuclei}
In the previous section, we found that UHECR nuclei can survive for external shocks. For internal shocks, the survival is difficult for Swift J1644+57-like TDEs, but lower-luminosity TDEs allow UHECR nuclei to escape without significant disintegration. In this section, for simplicity, we assume that UHECR nuclei survive and escape into intergalactic space and see consequences of different composition models~\footnote{This can be a reasonable approximation if the luminosity function has a steep index, i.e., the luminosity density is dominated by lower-luminosity TDEs.}. 

We assume that UHECR nuclei follow a power-law distribution with an exponential cutoff as
\begin{equation}
\frac{dN_{A^\prime}}{dE^\prime} = f_{A^\prime} N_0 \left( \frac{E^\prime}{Z E_0}\right)^{-s_{\rm esc}} {\rm exp} \left( -\frac{E^\prime}{Z E_{p, {\rm max}}^\prime}\right),
\end{equation}
where $f_{A^\prime}$ is the number fraction of different particles with mass $A^\prime$ at the same rigidity, $N_0$ is determined by the total energy per TDE event (see Appendix~\ref{Appendix_B}), $s_{\rm esc}$ is the spectral index of ejected UHECR nuclei, and $ Z E^\prime_{p, \rm max}$ is the maximum acceleration energy for particles with charge number $Z$. We assume the minimum particle energy $E^\prime_{A,\rm min}=\Gamma^2 Am_p c^2$. Strictly speaking, this is justified in the forward shock model but our main conclusion does not change by this assumption.

\subsection{Injection spectrum}
In order for UHECR nuclei to originate from stellar material (whether it is a MS star or WD), CR injections should occur inside jets or winds, which involve internal shocks and an external reverse shock. Details are highly uncertain but there are various possibilities. 

The first possibility is to rely on the shock acceleration at internal shocks. In the diffusive shock acceleration mechanism, accelerated particles have a power-law distribution, $dN/dE \propto E^{-s_{\rm acc}}$, with a typical spectral index $s_{\rm acc} \sim 2.2$ in the ultrarelativistic limit for the small-angle scattering approximation \cite{Kirk:2000yh, Achterberg:2001rx, Keshet:2004ch}.
However, the spectral index can be affected by the deflection angle, the ratio of the scattering mean free path to the particle gyroradius and the orientation of the magnetic field to the shock normal. In the large-angle scattering case, where magnetic fluctuations are sufficiently large, particles can gain significant energy in the single scattering and may lead to a harder spectrum \cite{Kato:2000hd, Aoi:2007aj, Baring:2010tn}. 
Such a possibility has been discussed to explain a hard spectrum of blazars \cite{Baring:2009iwa, Baring:2010tn}. 
In addition, the spectrum of escaping CRs does not have to be the same as that of accelerated CRs (e.g.,~\cite{Ohira:2009rd,Caprioli:2008sr,Katz:2010tv}; see also \cite{Fang:2017zjf,Globus:2014fka,Baerwald:2013pu}). If CRs could be confined for a long time after the dynamical time, CRs will lose their energies during their diffusive escape, so that the spectrum of escaping CRs is harder. Although details depend on flow dynamics and magnetic field evolution, one can assume that UHECRs escaping from internal shocks may have a small index such as $s_{\rm esc} < 2$. For an expanding outflow, one of the most conservative possibilities is to invoke the direct escape of CRs (e.g.,~\cite{Baerwald:2013pu}), which leads to $s_{\rm esc}=s_{\rm acc}-1$. For $s_{\rm acc}\sim2$, one may expect $s_{\rm esc}\sim1$.

The second possibility is to invoke a two-step acceleration mechanism via stochastic acceleration in the downstream of external shocks or possible reverse shock acceleration. The downstream may be highly turbulent and mixed around the contact discontinuity due to Rayleigh-Taylor instabilities \cite{Levinson:2009du}, and a fraction of energized nuclei from jets can be used for further acceleration to ultrahigh energies at the external forward shock \cite{Dermer:2000gu}. If heavy nuclei from jets can be used for CRs and the maximum energy is limited by the amplified magnetic field, accelerated CRs escape from a relativistic decelerating blast wave. The escaping spectrum can be harder in some cases \cite{Katz:2010tv} (see Ref.~\cite{Ohira:2009rd} for discussion on supernova remnants and AGN cocoon shocks). 
The evolution of total energy follows ${\mathcal E}_{\rm CR} \propto r^{-\alpha_{\mathcal E}}$, the evolution of external medium density is $\varrho \propto r^{-\alpha_\varrho}$, and the evolution of shocked fluid bulk Lorentz factor is $\Gamma \propto r^{-\alpha_\Gamma}$. The minimum energy of accelerated particles can be estimated as $E_{A, \rm min} \simeq \Gamma^2 A m_p c^2 \propto r^{-\alpha_{\rm min}}$, and the maximum particle energy is $E_{A, \rm max} \simeq Z e B r \propto r^{-\alpha_{\rm max}}$, as we discussed in the previous section. In the adiabatic expansion case, we have $\alpha_{\rm min} = 2 \alpha_\Gamma$, $\alpha_{\rm max} = \alpha_\Gamma + (1/2) \alpha_\varrho - 1$, and $\alpha_{\mathcal E} = 2 \alpha_\Gamma + \alpha_\varrho - 3$. We assume the accelerated CRs spectrum $\propto E^{-s_{\rm acc}}$, and have spectrum $\propto E^{-s_{\rm esc}}$ after escape from the acceleration site; Ref.~\cite{Katz:2010tv} derived a simple analytic relation between $s_{\rm esc}$ and $s_{\rm acc}$ as $s_{\rm esc} = s_{\rm acc} - (\alpha_{\rm min} (s_{\rm acc} - 2) + \alpha_{\mathcal E}) / \alpha_{\rm max}$ with the assumption that the number of ejected UHECRs in an energy interval is the same as that of the CRs at radius $r$, where the maximum acceleration energy is achieved. Let us adopt a typical value of power-law index $s_{\rm acc} \sim 2.2$. In the adiabatic expansion scenario, we have $\alpha_{\mathcal E} = 0$. If the CNM density is constant, then we have $\alpha_\varrho = 0$. The escaping particle spectral index can be calculated as $s_{\rm esc} = -5s_{\rm acc} + 12 = 1$. However, if the CNM density decreases with radius as $\varrho \propto r^{-\alpha_\varrho}$, we have $s_{\rm esc} = 1.4$ for $\alpha_\varrho = 1$ and $s_{\rm esc} = 1.8$ for $\alpha_\varrho = 2$. The CNM gas density in the galaxy nuclear region is still unclear. It can originate from the stellar winds, and their profile depends on the detailed distribution of stars and star formation history in the galaxy nuclear region \cite{Generozov:2015wra, Generozov:2016oon}. We assume that the CNM density is constant in our analysis, which is reasonable for galaxy cores \cite{Generozov:2015wra}. 

Although the above arguments are rather speculative, it is possible to expect a hard spectral index for escaping CRs, and we use $s_{\rm esc}\sim1$ in the jetted TDE scenario. In addition, we assume the maximum acceleration energy $E_{A,\rm max}$ according to the rigidity dependence, which is valid in the internal shock model for relatively low-luminosity TDEs and the external forward shock model.

\subsection{Composition model}
\subsubsection{MS-SMBH tidal disruptions}
\begin{figure}
\includegraphics[width=\linewidth]{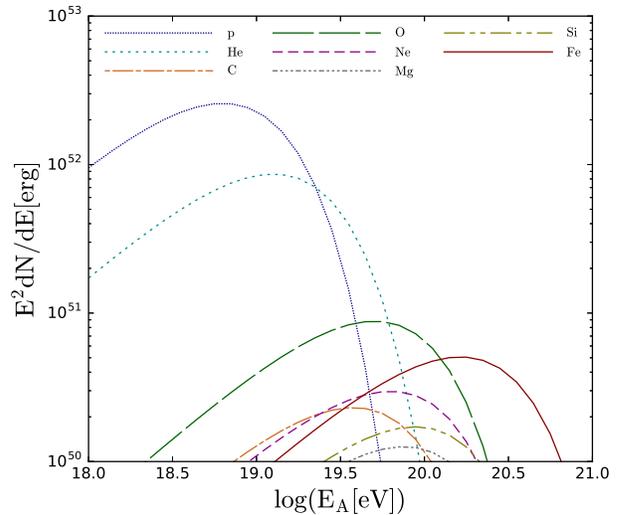}
\caption{CR injection spectra for model \RN{1}: MS stars tidally disrupted by SMBHs. The mass fraction is $X_{\rm H} = 73.9\%$, $X_{\rm He} = 24.7\%$, $X_{\rm C} = 0.22\%$, $X_{\rm N} = 0.07\%$, $X_{\rm O} = 0.63\%$, $X_{\rm Ne} = 0.17\%$, $X_{\rm Mg} = 0.06\%$, $X_{\rm Si} = 0.07\%$, and $X_{\rm Fe} = 0.12\%$ , which are taken from Ref.~\cite{Lodders:2010kx}. The maximum energy is $Z E_{p, \rm max} = 2 \times 10^{19} Z \rm~eV$, and the spectral index is $s_{\rm esc} = 1$. The total CR injection energy is ${\mathcal E}_{\rm CR} = 10^{53} \ \rm erg$. \label{fig:Spectrum_source_MS}}
\end{figure}

In model \RN{1}, we consider a solar-type MS star disrupted by a SMBH \cite{Rees:1988bf}. The present-day solar composition for H, He and metals is $X = 0.793$, $Y = 0.2469$, and $Z = 0.0141$, and the most abundant heavy nuclei are $\rm O$, $\rm C$, $\rm Ne$, and $\rm Fe$ \cite{Lodders:2010kx}.
In Fig.~\ref{fig:Spectrum_source_MS}, we show the CR injection spectra for model  \RN{1}. The mass fractions $X_A$ for different chemical species at the same rigidity are similar to the MS star. We assume that the maximum acceleration energy is $ Z E_{p, \rm max} = 2 \times 10^{19} Z \ \rm~eV$. 

\subsubsection{WD-IMBH tidal disruptions}
One of the models for Swift J1644+57 is a WD tidally disrupted by a $10^4 M_\odot$ IMBH \cite{Burrows:2011dn, Bloom:2011xk, Zauderer:2011mf, Krolik:2011fb}. The radii of WDs are smaller than those of MS stars, so WDs should be tidally disrupted at smaller radii from the central BHs. We need the factor $\beta_g = \frac{R_t}{R_g} > 1$ to avoid WDs being swallowed by BHs, where $R_t$ is the tidal disruption radius and $R_g$ is the Schwarzschild radius. The upper limit on the BH mass is estimated to be $M_{\rm BH} \lesssim 2.6 \times 10^{5} \ {M_\odot} \left( \frac{R_{\rm WD}}{10^9 \ \rm cm}\right)^{3/2} \left( \frac{M_{\rm WD}}{0.6 M_\odot}\right)^{-\frac{1}{2}}$, where $R_{\rm WD}$ and $M_{\rm WD}$ are the radius and mass of WDs \cite{Luminet1989}. 
Most of the WDs are composed of carbon and oxygen as the result of helium burning in the core, and a fraction of WDs may contain neon and magnesium due to the ignition of carbon. The hydrogen rich envelope of WDs could be ejected due to helium thermal pulses in the asymptotic giant branch phase. 

\begin{figure}
\includegraphics[width=\linewidth]{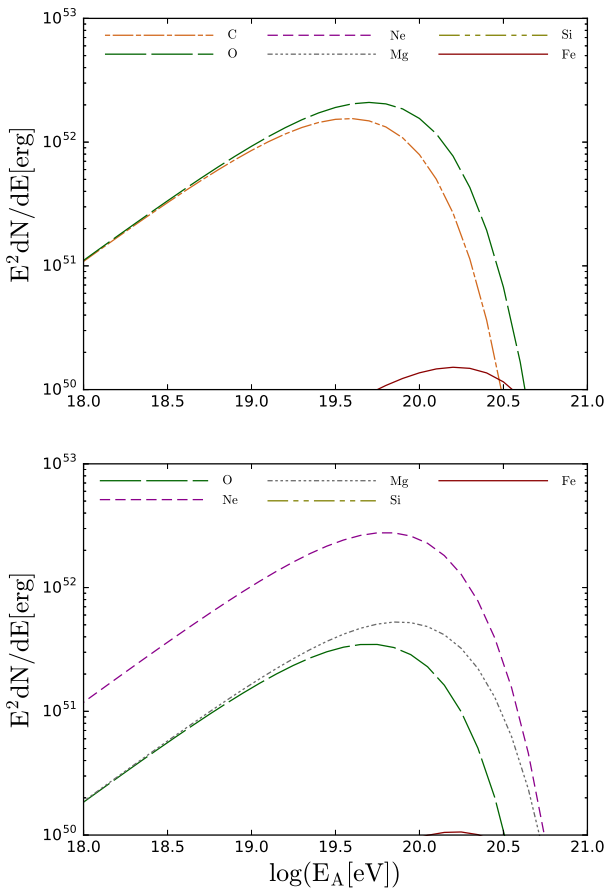}
\caption{ CR injection spectra for WDs disrupted by IMBHs. Upper panel: Model \RN{2}-1, CO-WDs with mass fraction $X_{\rm C} = 0.5$, $X_{\rm O} = 0.5$. Lower panel: Model \RN{2}-2, ONeMg-WDs with mass fraction $X_{\rm O} = 0.12$, $X_{\rm Ne} = 0.76$, $X_{\rm Mg} = 0.12$. The maximum energy is $Z E_{p, \rm max} = 6.3 \times 10^{18} Z \rm~eV$, and the spectral index is $s_{\rm esc} = 1$. The total CR injection energy is ${\mathcal E}_{\rm CR} = 10^{53} \ \rm erg$.  \label{fig:Spectrum_source_CO_ONeMg}}
\end{figure}

In model \RN{2}-1, we consider that tidally disrupted WDs are carbon-oxygen WDs (CO-WDs). CO-WDs are the end point for low and intermediate mass stars in the mass interval from $0.6 M_\odot$ to $6 M_\odot$ \cite{Antona1990, Weidemann2000}. The ratio of carbon to oxygen depends on the reaction ${}^{12}{\rm C}(\alpha, \gamma){}^{16}{\rm O}$, which is unresolved yet in nuclear physics \cite{Horowitz2010, Fields2016}. We assume that CO-WDs have a roughly equal mass fraction for carbon and oxygen $X_{\rm C} = 0.5$, $X_{\rm O} = 0.5$.  Stars with a mass in the range of $\sim 8 M_\odot$ to $\sim 12 M_\odot$ may evolve to form oxygen-neon-magnesium WDs (ONeMg-WDs) \cite{Nomoto1984, Iben1985}. In this case, the burning of carbon will not lead to explosion or collapse. In model \RN{2}-2, we consider that ONeMg-WDs with a mass fraction of $X_{\rm O} = 0.12$, $X_{\rm Ne} = 0.76$, and $X_{\rm Mg} = 0.12$ are tidally disrupted \cite{Isern1991}. In Fig.~\ref{fig:Spectrum_source_CO_ONeMg}, we show the CR injection spectra for model \RN{2}-1 (upper panel) and model \RN{2}-2 (lower panel). 

\subsubsection{WD-IMBH with ignition}
When a WD approaches a massive BH, the tidal compression and relativistic effects can enhance the WD central density and could trigger explosive nuclear burning \cite{Luminet1989, Wilson:2003iq, Dearborn2005, Rosswog:2007ke, Rosswog:2008ie, Haas:2012bk, Sell:2015rfa, MacLeod:2015jma, Tanikawa:2017skm}. This kind of ignition has also been suggested as an alternative scenario for type \RN{1}a supernovae \cite{Wilson:2003iq}. In this case, a fraction of nuclear explosive matter can be accreted into the center BH and form an accreting flow \cite{Rosswog:2008ie}. 

However, ignition and associated nucleosynthesis of heavy nuclei have been questioned by more dedicated simulations, and details are still under debate \cite{Tanikawa:2017skm}. Also, the rate of such events is uncertain and may be much smaller. On the other hand, it has been argued that the composition of such TDEs has been considered to explain UHECRs \cite{AlvesBatista:2017shr}, so we also consider this scenario for completeness. 

\begin{figure}
\includegraphics[width=\linewidth]{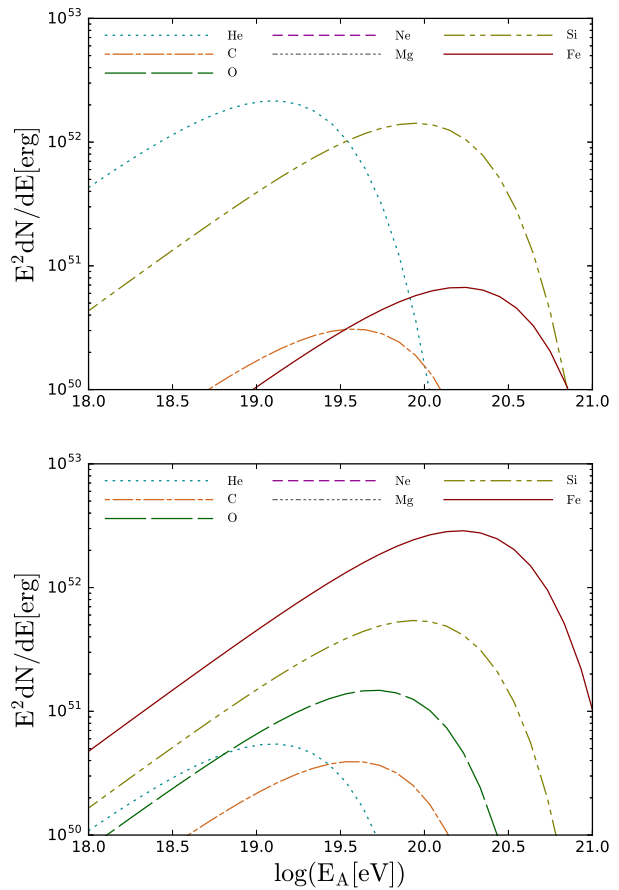}
\caption{CR injection spectra for WDs disrupted by an IMBH. Upper panel: Model \RN{3}-1, $0.2 M_\odot$ helium WDs disrupted by $10^3 M_\odot$ IMBHs. The mass fraction is $X_{\rm He} = 77.6\%$, $X_{\rm C} = 0.37\%$, $X_{\rm Si} = 7.3\%$, and $X_{\rm Fe} = 0.2\%$. 
Lower panel: Model \RN{3}-2, $1.2 M_\odot$ CO-WDs disrupted by $500 M_\odot$ IMBHs \cite{Rosswog:2008ie}. The mass fraction is $X_{\rm He} = 15.25\%$, $X_{\rm C} = 3.7\%$, $X_{\rm O} = 10.3\%$, $X_{\rm Ne} = 0.3\%$, $X_{\rm Mg} = 0.37\%$, $X_{\rm Si} = 21.7\%$, and $X_{\rm Fe} = 66.7\%$. The maximum energy is $Z E_{p, \rm max} = 6.3  \times 10^{18} Z\rm~eV$, and the spectral index is $s_{\rm esc}= 1$. The total CR injection energy is ${\mathcal E}_{\rm CR} = 10^{53} \ \rm erg$. \label{fig:Spectrum_source_Rosswog}}
\end{figure}

In models \RN{3}-1 and \RN{3}-2, we adopt the numerical simulation results from Ref.~\cite{Rosswog:2008ie}. Model \RN{3}-1 is the case with a $0.2 M_\odot$ helium WD passing a $10^3 M_\odot$ BH and model \RN{3}-2 corresponds to a CO-WD ($1.2 M_\odot$) approaching a $500 M_\odot$ BH . In Fig.~\ref{fig:Spectrum_source_Rosswog}, we show the CR injection spectra for model \RN{3}-1 (upper panel) and model \RN{3}-2 (lower panel).

\subsection{Propagation in intergalactic space}
We calculate the propagation of UHECR nuclei, using the public code CRPropa 3 \cite{Armengaud:2006fx, Batista:2016yrx}. The main energy-loss process for UHECR nuclei during propagation is photodisintegration due to CMB photons and extragalactic background light (EBL) photons. The EBL photons have a larger effect on the propagation of lower-energy intermediate mass nuclei ($\sim 10^{19} \ \rm~eV$) \cite{Allard:2011aa, Aab:2016zth}. In this work, we adopt a semianalytic EBL model derived by Ref.~\cite{Gilmore2012}. The details of the simulation are shown in Appendix \ref{Appendix_B}.  The observed UHECR spectrum has been accurately measured by Auger \cite{Aab:2015bza} and TA\cite{Fukushima:2015bza}. 
We use an empirical model, which describes the distribution of $X_{\rm max}$ using the generalized Gumbel function ${\mathcal G}(X_{\rm max})$ to get the mean value of the depth of shower maximum $\langle X_{\rm max} \rangle$ and $\sigma(X_{\rm max})$ \cite{DeDomenico:2013wwa}. In this work, we adopt the EPOS-LHC hadronic interaction model \cite{Werner:2010ss, Aab:2014kda}. 

\begin{figure}
\includegraphics[width=\linewidth]{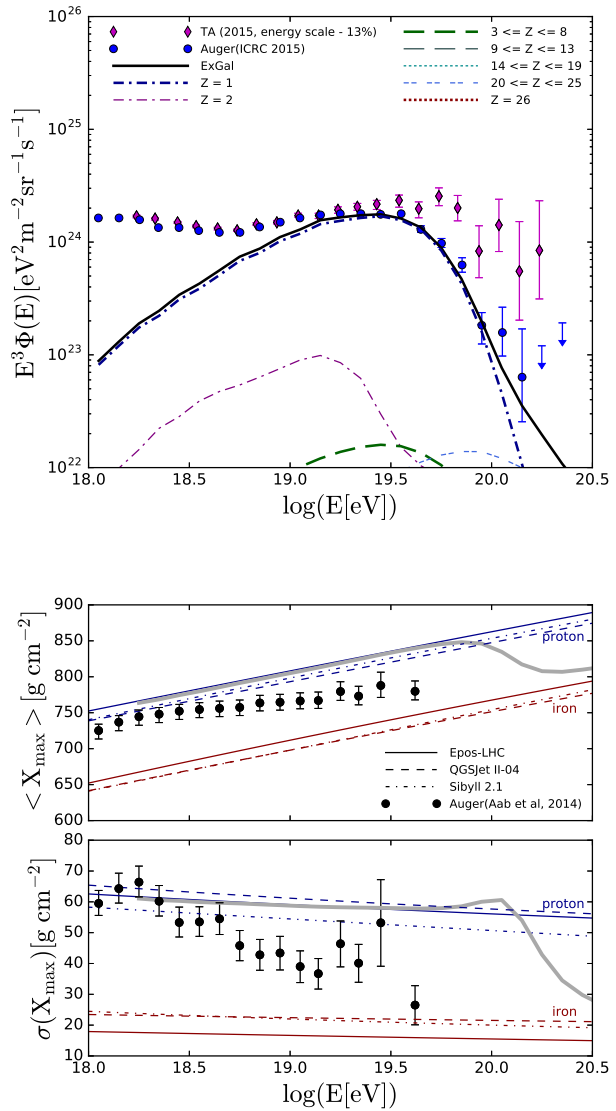}
\caption{ Model \RN{1}: MS stars with the solar composition. We use a maximum proton energy of $E_{p, \rm max} = 2 \times 10^{19} \ \rm~eV$ and spectral index of $s_{\rm esc}= 1$. \label{fig:Spectrum_MS}}
\end{figure}

In Fig.~\ref{fig:Spectrum_MS}, we show the results derived from model \RN{1}. This model can fit the observed UHECR spectrum measured by Auger, but failed to fit $\langle X_{\rm max} \rangle$ and $\sigma(X_{\rm max})$; the final spectrum is proton dominated in nearly the entire energy range ($< 10^{20} \ \rm~eV$). Although the situation is still under debate due to significant uncertainties in hadronic interaction models, this proton dominated scenario seems in strong tension with Auger results \cite{Aab:2014aea}.

\begin{figure}
\includegraphics[width=\linewidth]{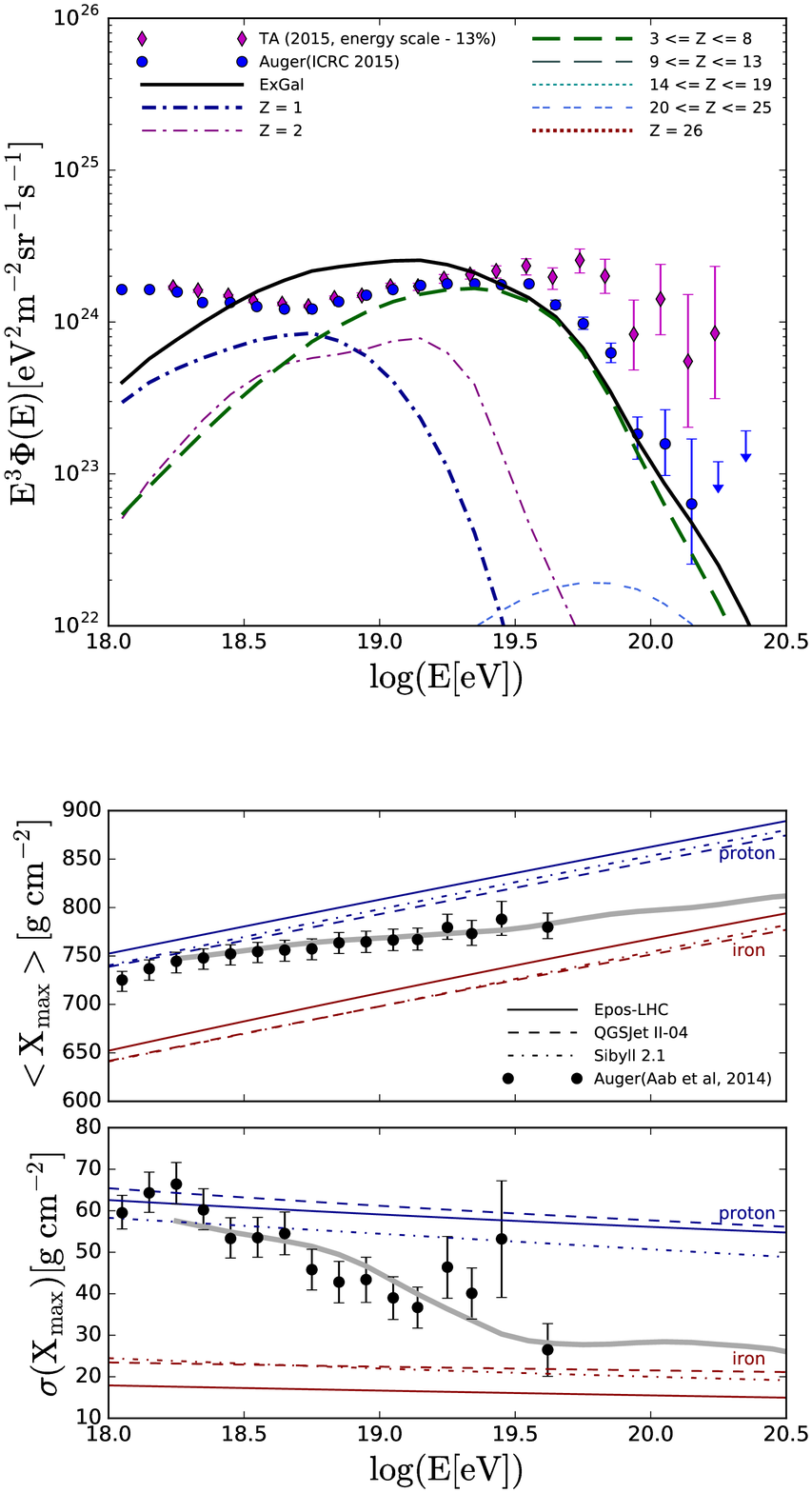}
\caption{ Model \RN{2}-1: CO-WDs with an initial mass composition, $X_{\rm C} = 0.5$, $X_{\rm O} = 0.5$. We use a maximum proton energy of $E_{p, \rm max} = 6.3 \times 10^{18} \rm~eV$ and spectral index of $s_{\rm esc} = 1$. \label{fig:Spectrum_CO}}
\end{figure}

\begin{figure}
\includegraphics[width=\linewidth]{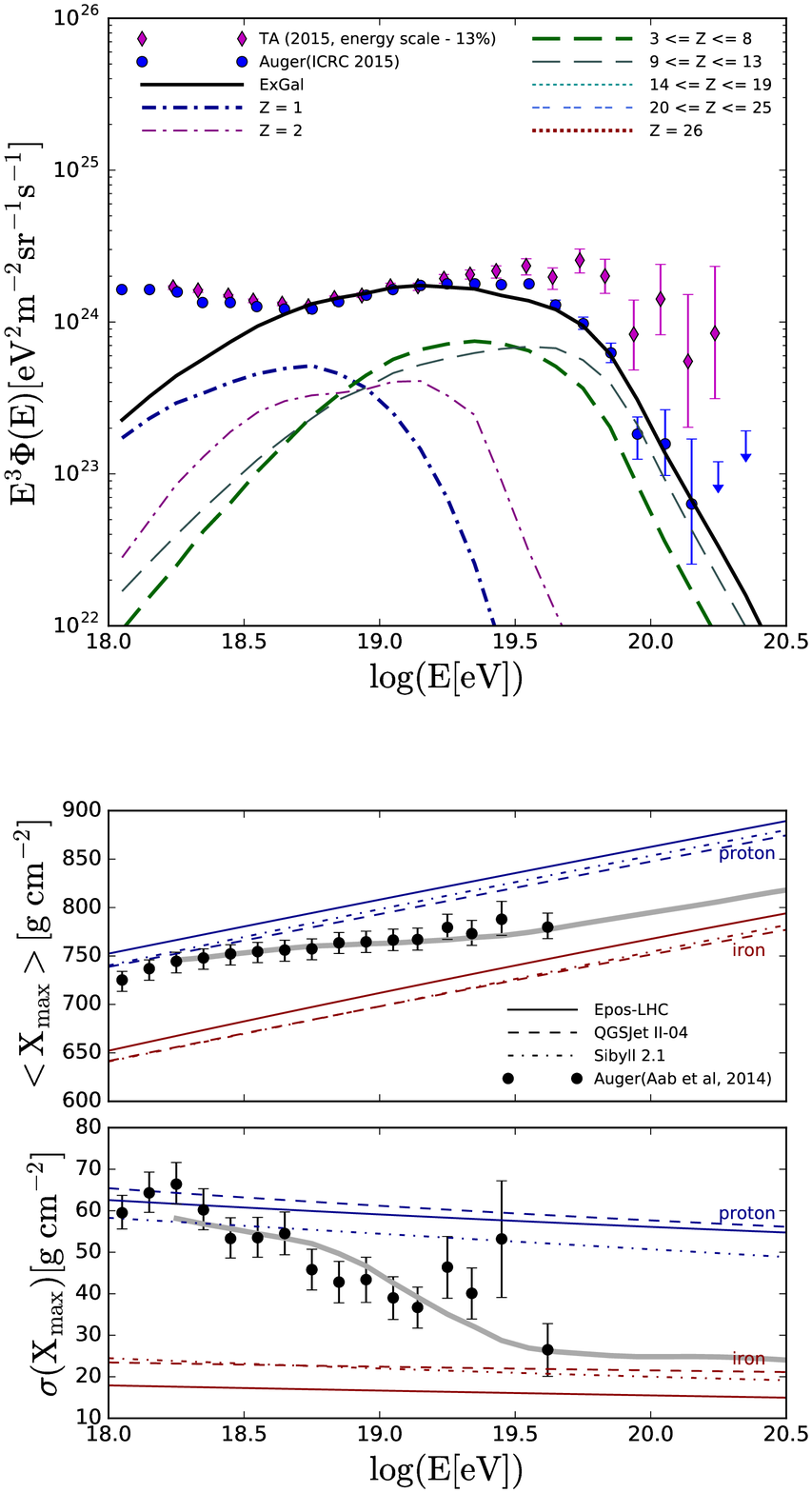}
\caption{ Model \RN{2}-2: ONeMg-WDs with an initial mass composition $X_{\rm O} = 0.12$, $X_{\rm Ne} = 0.76$, $X_{\rm Mg} = 0.12$. We use a maximum proton energy of $E_{p, \rm max} = 6.3 \times 10^{18}\rm~eV$ and spectral index of $s_{\rm esc} = 1$.\label{fig:Spectrum_ONeMg}}
\end{figure}

In Fig.~\ref{fig:Spectrum_CO}, we show the results from model \RN{2}-1, where most UHECRs are carbon and oxygen. The energy-loss lengths of C and O nuclei are very short, so most of the observed C and O nuclei should come from nearby sources. There should be a large fraction of secondary protons and helium generated during the propagation. We find that model \RN{2}-1 can fit the UHECR spectrum measured by Auger reasonably well, except the spectrum becomes softer in the "ankle" region ($\sim 10^{18.6} \rm~eV$). In Fig.~\ref{fig:Spectrum_ONeMg}, we show the results from model \RN{2}-2. In this model, UHECRs are mostly heavier nuclei, Ne and Mg, which is expected when WDs have higher mass such as ONeMg WDs. We find that the final spectrum can be fitted very well in this scenario. The data of $\langle X_{\rm max} \rangle$ and $\sigma(X_{\rm max})$ in both models are consistent with Auger results. Our results are consistent with their UHECR composition that becomes heavier with increasing energy, and where intermediate mass particles dominate in the high-energy range ($E > 10^{19} \ \rm~eV$).

\begin{figure}
\includegraphics[width=\linewidth]{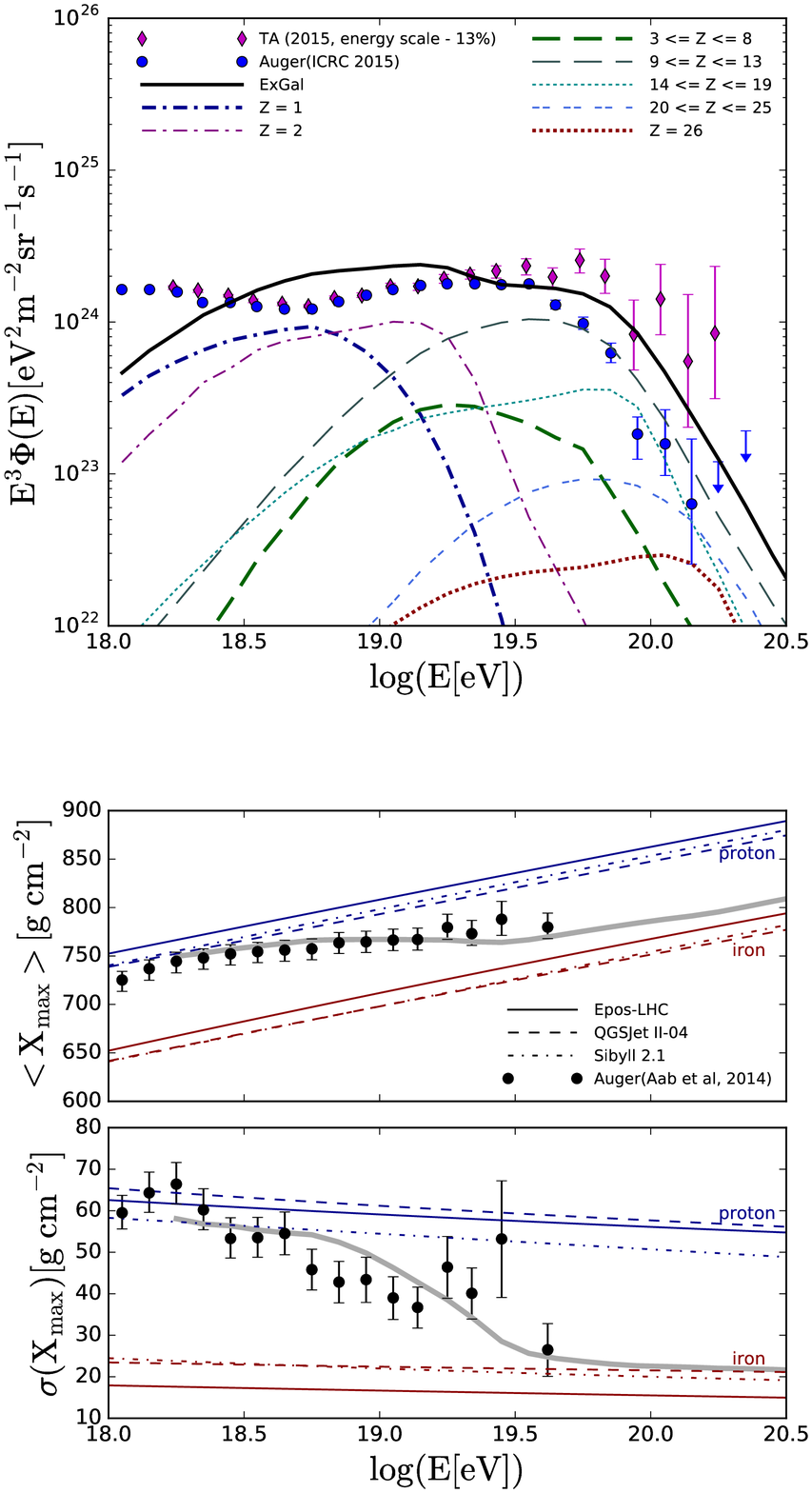}
\caption{ Model \RN{3}-1: $0.2 M_\odot$ (ignited) helium WDs disrupted by $10^3 M_\odot$ IMBHs. We use a maximum proton energy of $E_{p, \rm max} = 6.3 \times 10^{18}\rm~eV$ and spectral index of $s_{\rm esc} = 1$.      \label{fig:Spectrum_Rosswog_1}}
\end{figure}

\begin{figure}
\includegraphics[width=\linewidth]{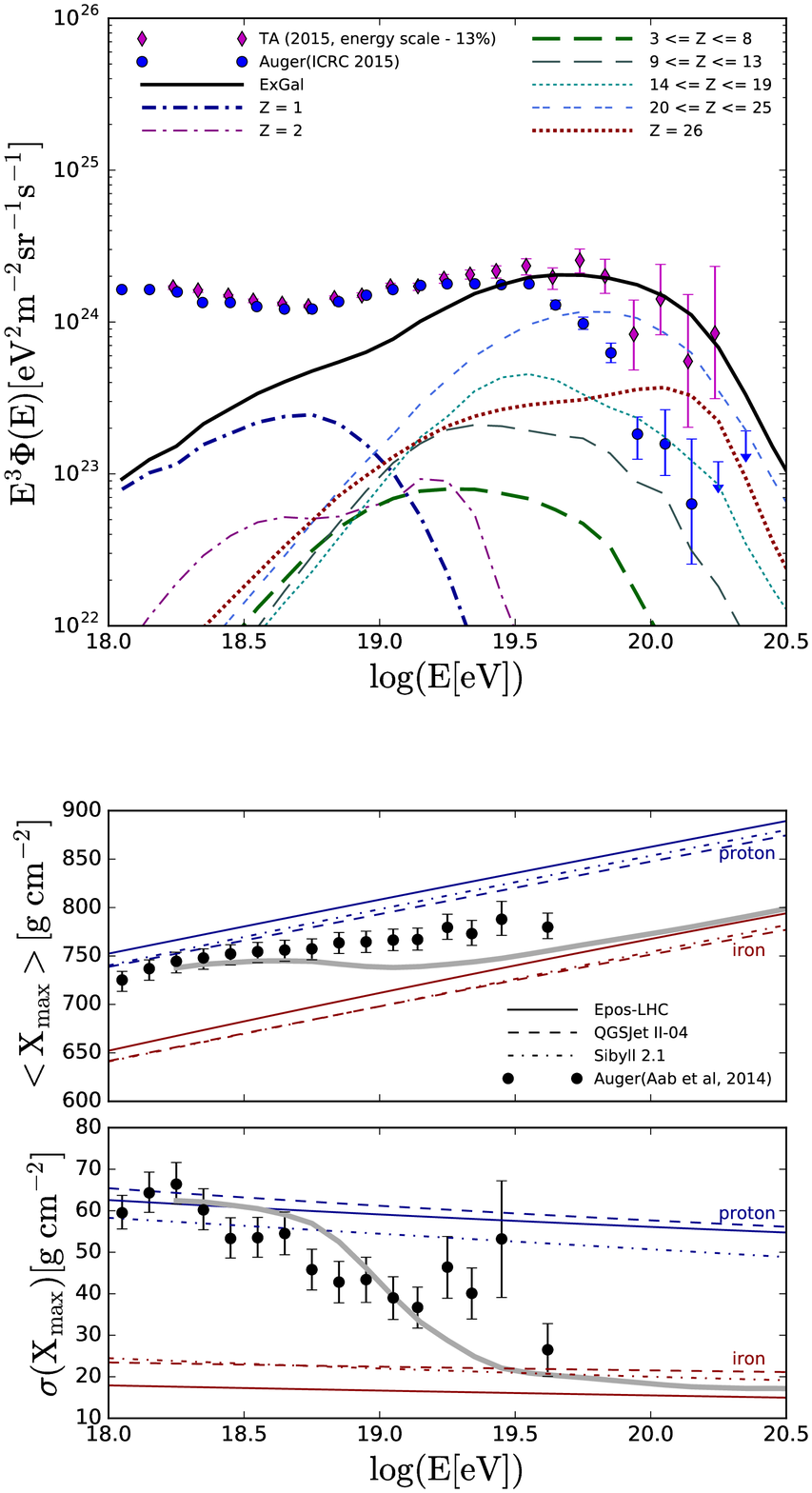}
\caption{Model \RN{3}-2: $1.2 M_\odot$ (ignited) CO-WDs disrupted by $500 M_\odot$ IMBHs. We use a maximum proton energy of $E_{p, \rm max} = 6.3 \times 10^{18}\rm~eV$ and spectral index of $s_{\rm esc} = 1$.\label{fig:Spectrum_Rosswog_2}}
\end{figure}

For the tidal disruption of lower-mass WDs (helium WDs or CO-WDs) by massive BHs, we also consider the enhancement of heavier nuclei through explosive nuclear reactions. Our results are shown in Fig.~\ref{fig:Spectrum_Rosswog_1} for model \RN{3}-1 and Fig.~\ref{fig:Spectrum_Rosswog_2} for model \RN{3}-2. UHECRs in model \RN{3}-1 are predominantly He, Si and Fe. This model can fit the UHECR spectrum reasonably in the lower energy range, but the cutoff energy ($\sim 10^{20} \ \rm~eV$) is more consistent with TA results and higher than Auger results. Compared to model \RN{3}-1, model \RN{3}-2 has a large fraction of iron group nuclei in UHECRs which predict an even higher cutoff energy. We find that model \RN{3}-1 is consistent with $\langle X_{\rm max} \rangle$ and $\sigma(X_{\rm max})$ measured by Auger, while model \RN{3}-2 has a large fraction of heavier nuclei compared to the observed composition data in the higher-energy range $(> 10^{19.2}\rm~eV)$. 

\section{\label{sec:four}Discussion and summary}
This work consists of two parts. 
In the first part, we examined the production of UHECRs in TDEs accompanied by relativistic jets. In the internal shock model, we found that CRs can be accelerated to ultrahigh energies. However, it is difficult for UHECR nuclei to survive in luminous TDE jets such as Swift J1644+57, while the survival is allowed for less powerful TDE jets. In the reverse and forward shock model, we can expect the production and survival of UHECRs. We also considered the wind model, where a nonrelativistic outflow interacts with the CNM. We found that in this scenario CRs can be accelerated only up to $\sim 10^{15}~Z \rm~eV$. 

In the second part, we examined different composition models for TDEs. Motivated by the composition data by Auger, we assumed that the injected UHECR nuclei mainly originate from tidally disrupted stars. Although we discussed several possibilities, this should be justified by more detailed work on CR acceleration and escape processes. We consider both MS-SMBH and WD-IMBH tidal disruptions. In the MS-SMBH TDE scenario (model \RN{1}), the injected UHECRs have a solarlike composition. The UHECR spectrum can be fitted, but a proton dominated composition is expected in the nearly the entire energy range. In the WD-IMBH TDE scenario, model \RN{2}-1 (CO-WDs) can give a poor fit to the UHECR spectrum but $\langle X_{\rm max} \rangle$ and $\sigma(X_{\rm max})$ can be reasonably accounted for. We found that it is difficult to fit the spectrum and composition simultaneously even if we try a variety of parameter sets, such as higher maximum energy and/or steeper ejection spectra. The main reason is that the attenuation lengths of UHECR carbon or oxygen nuclei are lower than protons or iron nuclei, so most of them will be depleted into secondary protons before reaching Earth \cite{Kotera:2011cp}. For ONeMg-WDs, we found that the results are more consistent with Auger data. However, the rate density of ONeMg-WDs is expected to be lower than that of CO-WDs ($\sim 1/30$), if we assume the Salpeter initial mass function \cite{Livio1994, Gil-PonsP.2003, Denissenkov2014} (see also Appendix C). The uncertainties of TDEs redshift evolution do not affect our results very much because most UHECRs mainly come from nearby sources within the GZK horizon ($\sim 100 \ \rm Mpc$) \cite{Kotera:2011cp}.

We also considered special cases, where nucleosynthesis is triggered by an IMBH.  Our results show that model \RN{3}-1 can marginally fit the observed spectrum, and consistent with the composition data, while model \RN{3}-2 has too many iron group nuclei, making it difficult to reconcile with the Auger data. We also caution that the final composition is sensitive to details of the parameters, such as WD and BH masses \cite{Rosswog:2008ie, MacLeod:2015jma, Kawana:2017pnq}. Reference~\cite{Tanikawa:2017skm} performed 3-D smoothed particle hydrodynamics simulations to study the explosive nuclear burning of WD-IMBHs, considering both the adiabatic compression and shock wave generation. They found that the final mass fractions are sensitive to the number of simulation particles, and ignitions occur for low-resolution simulations. 

Secondary gamma-ray and neutrino signals are of interest to test the model. TDEs have been discussed as high-energy neutrino sources~\cite{2008AIPC.1065..201M, MT09, Wang:2011ip, Wang:2015mmh, Dai:2016gtz, Senno:2016bso, Lunardini:2016xwi}. Neutrino production is expected to be efficient for luminous TDEs such as Swift J1644+57, while the neutrino production efficiency should be lower for low-luminous TDEs, allowing nucleus survival \cite{Murase:2008mr, Murase:2010va}. We also estimate the cosmogenic neutrino flux according to model \RN{2}-2 (ONeMg-WDs), and we find $E_\nu^2 \Phi_\nu \sim  {10}^{-10} \ \rm GeV \ cm^{-2} \ sr^{-1} s^{-1}$. However, this would be too conservative since luminous TDEs also contribute to the neutrino flux. 
High-energy gamma rays can be produced via neutron pion decay, photodeexcitation, and Bethe-Heitler processes. The survival of UHECR nuclei implies that gamma rays can escape from the sources without efficient two-photon annihilation inside the sources \cite{Murase:2008mr, Murase:2010va}. 

Perhaps, one could expect the emission of gravitational waves from WD-IMBH tidal disruptions~\cite{Kobayashi:2004py, Rosswog:2008ie, Haas:2012bk}, and TDEs can be interesting targets for multimessenger astronomy.

\medskip
\begin{acknowledgments}
We thank Shigeo S. Kimura and Kumiko Kotera for helpful discussions. 
The work of K.M. is supported by Alfred P. Sloan Foundation and the U.S. National Science Foundation (NSF) under grants NSF Grant No. PHY-1620777. It is partially supported by the National Natural Science Foundation of China under Grant No. 11273005 and the National Basic Research Program (973 Program) of China under Grant No. 2014CB845800 (Z.L.). B.T.Z. is supported by the China Scholarship Council (CSC) to conduct research at Penn State University.
\end{acknowledgments}

\appendix

\section{\label{Appendix_A} Inelasticity of photodisintegration process}
The giant dipole resonance (GDR) is the most important disintegration process at low energies, and the threshold energy is $\sim 8 \ \rm MeV$ \cite{Rachen1996, Khan:2004nd, Allard:2005ha}. The quasideutron (QD) process dominates when the NRF photon energy is larger than $30 \ \rm MeV$, and lower than the photopion production threshold energy. In this energy range, the wavelength of projectile photons is comparable with the size of nuclear, leading to the ejection of nucleon pair and additional nucleons. When the NRF photon energy is larger than $\sim 150 \ \rm MeV$, baryonic resonances (BR) play a dominant role in the photodisintegration process. At very high energies $\sim 1 \ \rm GeV$, a nucleus can be broken into many fragments of much lower energies via photofragmentation (PF).

In this work, the photodisintegration cross section for UHECR nuclei is taken from CRPropa 3 \cite{Armengaud:2006fx, Batista:2016yrx}. The cross section for nuclei with mass numbers of $A \geq 12$ and NRF photon energy in the range $0.2 - 200 \ \rm MeV$ can be derived using the nuclear reaction code TALYS \cite{Koning2007}. 
At higher energies, for simplicity, we use the relation that the cross section is proportional to nuclear mass, which can be written as $\sigma_{A\gamma}(\bar{\varepsilon}) = A \sigma_{p \gamma}(\bar{\varepsilon})$ (see Fig.~\ref{fig:crosssection}). The photomeson production cross section for protons is derived using the numerical code Geant4 \cite{Agostinelli:2002hh, Murase:2008mr}. 

In the photodisintegration process, a parent nucleus loses energy through the ejection of one or more nucleons. We assume that the Lorentz factor is constant during the interaction; then the relative energy loss can be estimated as \cite{Rachen1996}
\begin{equation}
\kappa_{A\gamma}(\bar{\varepsilon}) \equiv \frac{\Delta E}{E} = \frac{\Delta N}{N},
\end{equation}
where $N$ is the total number of nucleons in the parent nuclei, and $\Delta N$ is the number of ejected nucleons in each channel. To derive the average energy loss, we consider the contribution from all the dominant channels,
\begin{equation}
\bar{\kappa}_{A\gamma} = \frac{1}{\sigma_{\rm tot}}\sum_i \sigma_i \frac{[\Delta N]_i}{N},
\end{equation}
where $\sigma_{\rm tot}$ is the total cross section, $\sigma_i$ is the cross section for the ith channel, and $[\Delta N]_i$ is the number of ejected nucleons in the ith channel. In the energy range beyond the photopion production threshold, for simplicity, the inelasticity is assumed to linearly decrease with increasing nuclear mass $\bar{\kappa}_{A\gamma}(\bar{\varepsilon}) = \bar{\kappa}_{p\gamma}(\bar{\varepsilon}) / A$.

\section{\label{Appendix_B} UHECRs propagation}
The observed CR flux for each component is described by the following formula \cite{Unger:2015laa, Batista:2016yrx}
\begin{equation}
\Phi_A = \sum_{A^\prime} \Phi_{A A^\prime} = \sum_{A^\prime} \frac{c}{4\pi}\frac{dn_{A A^\prime}}{dE},
\end{equation}
where $\Phi_{A A^\prime}$ and ${\rm d}n_{A A^\prime}(E)$ are the observed cosmic ray flux and number density of particles with mass $A$ generated from parent particles with mass $A^\prime$. ${\rm d}n_{A A^\prime}(E)$ can be estimated by considering the contribution as a function of redshift
\begin{eqnarray}
dn_{A A^\prime}(E) &=& \int_{z_{\rm min}}^{z_{\rm max}} dz  \left| \frac{dt}{dz}\right| f_{\rm TDE}(z)\rho_0 \nonumber \\ &\times& \int_{E^\prime_{\rm min}}^{E^\prime_{\rm max}} dE^\prime \frac{dN_{A^\prime}}{dE^\prime}\eta_{A A^\prime}(E, E^\prime,z),
\end{eqnarray}
where
\begin{equation}
\frac{dt}{dz} = - \frac{1}{H_0 (1 + z)} \frac{1}{\sqrt{\Omega_\Lambda + \Omega_k(1 + z)^2 + \Omega_m(1 + z)^ 3}},
\end{equation}
and we use the redshift evolution of TDEs given by \cite{Sun:2015bda}
\begin{equation}
f_{\rm TDE}(z) = \left[ (1 + z)^{0.2 \eta} + \left( \frac{1+z}{1.43}\right)^{-3.2\eta} + \left( \frac{1+z}{2.66}\right)^{-7.0 \eta} \right]^{\frac{1}{\eta}},
\end{equation}
with $\eta \sim -2$. In this work, we adopt $z_{\rm min} = 0.0001$ and $z_{\rm max} = 2$. Note that the redshift evolution for TDEs is negative. $\rho_0$ is the local event rate of jetted TDEs \cite{Sun:2015bda}. Also, $\eta_{A A^\prime}(E, E^\prime, z)$ is the fraction of generated CRs with mass $A$ and energy $E$ from parent particles with mass $A^\prime$ and energy $E^\prime$. ${\rm d}N_{A^\prime}/{\rm d}E^\prime$ is the UHECR injection spectra per TDE.

The CR luminosity density can be estimated as
\begin{eqnarray}
Q_{\rm CR} &=& \sum_{A^\prime} \int_{E^\prime_{\rm min}}^{Z E^\prime_{p, \rm max}} dE^\prime E^\prime \frac{dN_{A^\prime}}{dE^\prime} \rho_0 \nonumber \\ &=& \sum_{A^\prime} f_{A^\prime} N_0 \rho_0 E_0^2 Z^2 \left(\frac{E^\prime_{p, \rm max}}{E_0} \right)^{2-s_{\rm esc}}  \nonumber \\ &\times & \Gamma(2-s_{\rm esc}, \frac{E^\prime_{\rm min}}{Z E^\prime_{p, \rm max}}),
\end{eqnarray}
where $\Gamma(2-s_{\rm esc}, \frac{E^\prime_{\rm min}}{Z E^\prime_{p, \rm max}})$ is the incomplete gamma function. We use the following formula to estimate the normalization parameter $N_0$:
\begin{equation}
{\mathcal E}_{\rm CR}^{\rm iso} = \sum_{A^\prime} \int_{E^\prime_{\rm min}}^{Z E^\prime_{p, {\rm max}}} dE^\prime E^\prime \frac{dN_{A^\prime}}{dE^\prime}.
\end{equation}
The CR energy per TDE (${\mathcal E}_{\rm CR}$) can be estimated through the relation ${\mathcal E}_{\rm CR}^{\rm iso} = \xi_{\rm CR} {\mathcal E}_{\rm rad}^{\rm iso}$, where $\xi_{\rm CR}$ is the cosmic ray loading factor. 

\section{\label{Appendix_C} Energy budget}
The event rate of WD-IMBH tidal disruptions is uncertain. IMBHs are believed to exist in dwarf galaxies or globular clusters. In dwarf galaxies, the event rate is estimated to be $R_{\rm TDE-DG} \sim 10 f_{\rm IMBH}^{\rm DG} \ \rm Gpc^{-3} \ yr^{-1}$, where $f_{\rm MBH}^{\rm DG}$ is the occupation fraction in dwarf galaxies \cite{MacLeod:2014mha}. In globular clusters, the event rate is $R_{\rm TDE-GC} \sim 50 f_{\rm MBH}^{\rm GC} \ \rm Gpc^{-3} \ yr^{-1}$, which is slightly higher than the event rate estimated for dwarf galaxies  \cite{Shcherbakov:2012zt}. 

We assume only a fraction of $f_{\rm jet} \sim 10\%$ WD-IMBH tidal disruptions have relativistic jets with the beaming factor $f_b \approx  \theta_j^2 /2 \sim 1/(2\Gamma_j^2)$, with a typical Lorenz factor of $\Gamma_j = 10$. The apparent rate density of WD-IMBH tidal disruptions is
\begin{equation}
\rho_{\rm WD-IMBH} \sim 0.06 \left(\frac{f_{\rm IMBH}}{1}\right) \left(\frac{f_{\rm jet}}{0.1}\right) \left(\frac{f_b}{0.01}\right) \ \rm Gpc^{-3} \ yr^{-1},
\end{equation}
with the assumption $f_{\rm IMBH} \simeq f_{\rm IMBH}^{\rm DG} \simeq f_{\rm IMBH}^{\rm GC} \simeq 1$. The event rate can be comparable to the event rate ($\rho_{\rm TDE}\simeq 0.03 \ \rm Gpc^{-3} \ yr^{-1}$) obtained from the detection of two jetted TDEs Swift J1655+57 and Swift J2058+05 \cite{Sun:2015bda}.

According to our simulation results, we need a total CR energy injection rate of $Q_{\rm CR} \approx 4.2 \times 10^{44} \ \rm erg \ Mpc^{-3} \ yr^{-1}$ in model \RN{2}-2, and the required CR loading factor is $\xi_{\rm CR} \sim 100 \ Q_{\rm CR, 44.6} {\rho_0}_{-10.2}^{-1} \ {\mathcal E}_{\rm rad, 53}^{-1}$. Note that the luminosity density itself is independent on the beaming factor as in the argument for GRBs. The absolute radiation energy is smaller than the isotropic-equivalent radiation energy by $f_b^-1$, whereas the true rate density is a factor of $f_b^-1$ larger than the apparent rate density of TDEs, of which jet points to us.

\bibliographystyle{apsrev}
\bibliography{bzhang}

\end{document}